# Molecular dynamics simulations of cRGD-conjugated PEGylated TiO$_2$ nanoparticles for targeted photodynamic therapy

Paulo Siani[a], Giulia Frigerio[a], Edoardo Donadoni[a], Cristiana Di Valentin[a,b,*]

[a]Dipartimento di Scienza dei Materiali, Università di Milano Bicocca,
via R. Cozzi 55, 20125 Milano Italy

[b]BioNanoMedicine Center NANOMIB, University of Milano-Bicocca, Italy

## Abstract

The conjugation of high-affinity cRGD-containing peptides is a promising approach in nanomedicine to efficiently reduce off-targeting effects and enhance the cellular uptake by integrin-overexpressing tumor cells. Herein we utilize atomistic molecular dynamics simulations to evaluate key structural-functional parameters of these targeting ligands for an effective binding activity towards $\alpha_V\beta_3$ integrins. An increasing number of cRGD ligands is conjugated to PEG chains grafted to highly curved TiO$_2$ nanoparticles to unveil the impact of cRGD density on the ligand's presentation, stability, and conformation in an explicit aqueous environment. We find that a low density leads to an optimal spatial presentation of cRGD ligands out of the "stealth" PEGylated layer around the nanosystem, favoring a straight upward orientation and spaced distribution of the targeting ligands in the bulk-water phase. On the contrary, high densities favor over-clustering of cRGD ligands, driven by a concerted mechanism of enhanced ligand-ligand interactions and reduced water accessibility over the ligand's molecular surface. These findings strongly suggest that the ligand density modulation is a key factor in the design of cRGD-targeting nanodevices to maximize their binding efficiency into over-expressed $\alpha_V\beta_3$ integrin receptors.

---

[*] Corresponding author: cristiana.divalentin@unimib.it

## 1. Introduction

Active targeting strategies, exploiting the biological interaction between ligands on the surface of nanoparticles (NPs) and the cell target, are receiving increasing attention because, in several cases, they have been found to increase the therapeutic efficacy with respect to passive targeting strategies based on physicochemical properties of the NPs and on the enhanced permeability and retention (EPR) effect of tumor cells [1].

Among cell targets, integrins are a promising class of receptors for two main reasons: first, they are overexpressed in many types of cancer cells [2] and tumor microvasculature and, secondly, they regulate the communication between cells and their microenvironment. Therefore, the surface functionalization of nanomaterials with integrin-specific ligands not only increases their binding affinity to cancer cells but also enhances the cellular uptake of the NPs thanks to the intracellular trafficking of integrins [3,4] or through more complex routes, such as phagocyte hitchhiking [5].

Recent advances in integrin-targeted therapeutics have proved that fine-tuning the density of high-affinity integrin ligands conjugated to the surface of nano-delivery systems, e.g., metal or metal oxide NPs, is a viable way to improve the clinical outcome [6]. On the one hand, it is intuitive to assume that increasing the availability and, therefore, the spatial density of targeting ligands on the NP might lead to a higher probability of binding events to targeted integrins, enhancing their internalization by tumor cells [7]. On the other hand, there is experimental evidence that increasing ligand density as a tunable parameter of a nano-delivery system might cause "off-targeting" effects impairing their targeting activity, e.g. off-targeting effect due to impaired stealth properties [8], over-crowding effects due to ligand-ligand interactions impairing their proper orientation for binding, over binding to cell receptors decreasing the uptake efficiency by target cells, among others [9].

From an empirical perspective, Abstiens et al. [10] have studied the interplay between the ligand density and the integrin-mediated cellular uptake of polymeric NPs decorated with cyclic tripeptide Arg-Gly-Asp molecules (c(RGDfK)). The authors found that NPs with short poly(ethylene glycol) (PEG) chains and high c(RGDfK)-conjugation density have superior efficiency in targeting $\alpha_v\beta_3$ integrins than NPs with longer PEG chains and lower c(RGDfK)-conjugation densities. In a more recent paper [8], the same authors have introduced a new concept of "steric shielding" for c(RGDfK)-conjugated polymeric NPs to avoid their off-targeting to endothelial integrins. They argued that, by fine-tuning the c(RGDfK) surface density and the length of PEG chains, undesired off-target accumulation may be prevented, while a satisfactory on-target efficacy may be maintained. In the same vein, Valencia et. al. [11] have found that copolymeric NPs composed of c(RGDyK)-conjugated PLGA(poly lactic-co-glycolic acid)-PEG chains show optimal ligand densities up to 50% for enhanced tumor cell uptake and minimum phagocytic clearance.

Theoretical predictions based on mathematical tumor models have shown that controlling the density of RGD-targeting ligands is a workable way to enhance their tumor-penetration efficiency with a stout effect on the integrin binding kinetics [12]. The authors have demonstrated that the addition of more than 3 cyclic RGD targeting ligands to poly(amidoamine) (PAMAM) dendrimers, by impairing the ligand's binding affinity to integrins, enhances the tumor penetration by PAMAM-RGD conjugates. Recently, Biscaglia et al. [13] have investigated, experimentally and by means of atomistic Molecular Dynamics (MD) simulations, the role of a cationic spacer on the presentation and orientation of c(RGDyK) ligands conjugated to PEGylated gold NPs. They found that c(RGDyK) ligands linked to PEGylated Au NPs through an oligolysine spacer have enhanced targeting activity towards integrins, whereas, in the absence of this spacer (when the cyclic RGD is directly bonded to PEG chains), the targeting efficacy is impaired.

Titanium dioxide ($TiO_2$) NPs are attracting increasing attention in nanomedicine because of their unique photocatalytic properties, high biocompatibility, low toxicity [14], low cost, and high chemical stability [15]. In particular, they are used in photodynamic therapy for cancer treatment [16–18], being excellent ROS (reactive oxygen species) generators under light irradiation, as drug delivery systems [19,20], and for cell imaging [21–23]. Coating $TiO_2$ NPs with PEG is a common strategy to elongate their circulation time [24,25]. On top of that, active targeting with RGD ligands could improve the selectivity of photodynamic therapy towards tumor cells.

In this work, we aim at elucidating, by means of fully atomistic MD simulations, the mechanism of how the density of c(RGDyK) (from now on cRGD) conjugation of PEGylated TiO$_2$ nanoparticles affects the ligand's conformation and presentation in an aqueous environment and, consequently, the ligand's targeting ability.

First, in Section 2, we present the computational methods and the details of the analysis performed in this work. Then, in Section 3.1, we present the separated studies of the single components of the nanodevice (cRGD, PEG chain, PEGylated-NP, etc.), which are useful not only to understand the actors involved but also to assess the reliability of the chosen setup and models, through comparison against experimental data and previous studies. Section 3.2 is devoted to the investigation of the whole targeting nanodevice: first, we examine the effect of cRGD density on the cRGD spatial distribution, also through a systematic comparison against available experimental data, and, secondly, we shed light on the role played by the cRGD density on the nanodevice's stabilization and diffusion in water. By analyzing the total interaction energy and its electrostatic and vdW components, we discuss the driving forces controlling the extent of exposure of cRGD ligands in the nanosystem and elucidate the underlying molecular mechanisms. Finally, we compare the resemblance of the various cRGD structures found in our simulations, at different ligand densities, to the X-ray atomic structure of a similar ligand in complex with a $\alpha_v\beta_3$ integrin for a fact-based assessment of the potential targeting ability. In the Conclusions section, we wrap up the paper addressing the implications of ligand density in the development of new nano-delivery systems for biomedical applications.

## *2. Computational Methods*

### *2.1 Simulation details*

#### 2.1.1 Single cRGD, PEG$_{500}$, and PEG$_{500}$-cRGD in water

The *Ligand Reader & Modeler* module of CHARMM-GUI [26,27] was employed to generate the three-dimensional structures of PEG$_{500}$ chains and their conjugation with cRGD ligands was done using the GaussView program [28]. Bonded and non-bonded parameters for cRGD, PEG$_{500}$ and PEG$_{500}$-cRGD were assigned by analogy using the CHARMM36m [29–31] and CGenFF [32–34] force fields (FFs). The partial atomic charges of cRGD and PEG$_{500}$ are reported in the Supporting Material (Fig. S1, Table

S1, Fig. S2, Table S2). The cRGD, PEG$_{500}$, and PEG$_{500}$-cRGD molecules were solvated by cubic boxes (56x56x56 Å$^3$, 60x60x60 Å$^3$, and 76x76x76 Å$^3$, respectively) filled up with mTIP3P [35–37] water molecules at the experimental density of about 0.99 g/cm$^3$ using the CHARMM-GUI *Solvation Builder* [26,38]. One Cl$^−$ counter-ion was added to the system containing the cRGD molecule to maintain the system's overall electroneutrality due to the positive net charge acquired by this ligand at physiological pH conditions. First, a minimization phase with 10000 steps using the conjugate gradient algorithm was performed, followed by a subsequent equilibration phase carried out in the NPT ensemble up to 50 ns and extended for more 50 ns of MD production for each system. The long-range solver Particle Mesh Ewald (PME) scheme [39] handled the electrostatic interactions with a cut-off distance of 12.0 Å in the real space with forces smoothly switched to zero after 10.0 Å. A Langevin thermostat with a damping coefficient of 1.0 ps$^{-1}$ and a semi-isotropic Nosé-Hoover Langevin-piston scheme with a piston period of 50 fs and a piston decay of 25 fs were coupled to the systems to keep the ensemble temperature and pressure constant at 303.15 K and 1 atm, respectively. Newton's equations of motion were solved in time using a 2 fs time step by the impulse-based Verlet-I/r-RESPA integrator [40]. A combination of SHAKE/RATTLE [41,42] and SETTLE [43] algorithms were employed to impose holonomic constraints on all covalent bonds involving hydrogen atoms. To validate the CHARMM-FF assignment of topologies and parameters for cRGD, PEG$_{500}$ and PEG$_{500}$-cRGD, we carried out, separately, MD simulations for each of these molecules using the NAMD2.13 code [44].

### 2.1.2 PEGylated and cRGD-conjugated PEGylated TiO$_2$ nanoparticle in water

To keep the PEGylated TiO$_2$ NP model consistent with our previous work [45], we used a bare 2.2 nm TiO$_2$ NP [46,47] as the substrate for the PEGylated TiO$_2$ NP model. In short, we grafted the bare 2.2 nm TiO$_2$ NP with 50 methyl-terminated PEG$_{500}$ chains, whose -OH terminal groups bind to 4-coordinate Ti atoms, which are the most reactive sites, and 5-coordinate Ti atoms on the TiO$_2$ NP surface. The resulting grafting density (σ) is equal to 0.02252 chain Å$^{-2}$, which was previously demonstrated to be near the transition from mushroom to brush polymer conformation [45]. Further details on the protocol employed to the TiO$_2$ PEGylation can be found elsewhere [45]. The above-cited PEGylated $TiO_2$ NP model was then conjugated with different cRGD ligand densities. Three cRGD ligand densities were studied: a lower one, with 5 out of 50 PEG chains conjugated with cRGD and corresponding to a degree of conjugation of 10% (0.2 ligand $nm^{-2}$); an intermediate one, with 10 out of 50 PEG chains conjugated with cRGD ligands reaching a degree of conjugation of 20% (0.5 ligand $nm^{-2}$); a higher one, with 25 out of 50 PEG chains conjugated with cRGD ligands reaching a degree of conjugation of 50% (1.1 ligand $nm^{-2}$). Here, the degree of conjugation is defined as the percentage of cRGD-$PEG_{500}$ chains out of the total $PEG_{500}$ chains, whether conjugated or not with cRGD. PACKMOL [48] was employed to center the systems in cubic boxes (100x100x100 $Å^3$ for the PEGylated $TiO_2$ NP and for the cRGD-conjugated PEGylated $TiO_2$ NPs with a degree of conjugation of 10% and 20% and 120x120x120 $Å^3$ for the NP with a degree of conjugation of 50%) which were filled up with mTIP3P [35–37] water molecules at the experimental density of about 0.99 $g/cm^3$. The CGenFF force field [32–34] was used to assign the partial atomic charges and Lennar-Jones (LJ) parameters for the $PEG_{500}$ and $PEG_{500}$-cRGD chains. The partial atomic charges and LJ (12,6) parameters for the $TiO_2$ NP were assigned according to the coordination number of titanium and oxygen atoms using an optimized version of the original Matsui-Akaogi FF [49] further refined by Brandt et al. [50]. This FF has been tested and validated by one of us for a $TiO_2$ NP model tethered with small organic molecules [51]. Lorentz-Berthelot rules were used to obtain the cross-term parameters for LJ interactions between unlike atoms. The long-range solver Particle-Particle Particle-Mesh (PPPM) handled the electrostatic interactions with a real-space cut-off of 10 Å and a threshold of $10^{-5}$ for the error tolerance in forces [52]. The Lennard-Jones (12,6) interactions were truncated with a 10 Å cut-off with a switching function applied beyond 8 Å. A minimization phase was carried out to minimize the total potential energy of the system, followed by an equilibration phase 200 ns long, keeping the $TiO_2$ NP core fixed at its QM-optimized geometry. In the MD production phase, the $TiO_2$ core was treated as an independent rigid body able to rotate and move in the system, as done in a previous work by some of us [53]. This approach holds the DFTB-optimized geometry and avoids any misshaping of the $TiO_2$ core during the MD simulation. The MD production phase explored 100 ns of the phase space in the NVT ensemble at 303.15 K held constant by the Nosé-Hoover thermostat with a damping coefficient of 0.1 $ps^{-1}$. Only the last 50 ns of production phase were considered for the analysis. Non-bonded interactions were treated according to the CHARMM potential energy functions [31,54,55], with a force switching distance of 10 Å and a real-space cut-off distance of 12 Å under Periodic Boundary Conditions. Newton's equations of motion were integrated in time using the Velocity-Verlet integrator with a timestep of 2.0 fs. The SHAKE algorithm [41] was used to impose holonomic constraints on all covalent bonds including hydrogen atoms. All MD simulations including the nanoparticle were carried out using the CHARMM implementation in the LAMMPS code (version 29 Oct 2020, http://www.lammps.org) [56].

## *2.2 Simulation analysis*

### 2.2.1 Cluster Analysis

To identify the ensemble of relevant conformations acquired by cRGD, either as a single molecule free to diffuse in water or conjugated to the $PEG_{500}$ chains, we carried out a cluster analysis using the GROMOS method [57]. To determine the cluster membership, we based the analysis only on heavy atoms of cRGD with a cutoff for RMSD of 2.0 Å.

### 2.2.2 Hydrogen-bonding

The hydrogen-bond formation through the MD trajectory was tracked using the Hydrogen Bonds tool implemented in VMD [58]. We counted as an H-bond formation when the following geometrical criteria were satisfied: (1) the distance between the H-donor and the H-acceptor heavy atoms was less than 3.5 Å; (2) the supplementary angle to the one formed between the H-donor and H-acceptor heavy atoms, having the hydrogen atom as the vertex, was less than 30 °.

### 2.2.3 Non-bonding interaction energy

To evaluate the relevant aspects of the intermolecular interactions established by the nanodevice components and their molecular surrounding during the MD simulation, we utilized the USER-TALLY package implemented in LAMMPS. The vdW and electrostatic components of intermolecular interactions are reported.

### 2.2.4 Structural parameters of polymers

To validate the CHARMM-based PEGylated $TiO_2$ NP model against existing literature data and further characterize the polymeric chain behavior in water, we analyzed some polymer structural descriptors, as detailed in the following.

To assess the average length of PEG chains, we measured the *end-to-end distance* ($\langle h^2 \rangle^{1/2}$), defined as the distance between the first and the last heavy atom of a polymeric chain. To estimate the approximated spherical volume occupied by the PEG chains tethered to the $TiO_2$ NP surface, we used the *radius of gyration* ($R_g$) analysis defined as follows:

$$R_g = \sqrt{\frac{1}{N}\left\langle \sum_{k=1}^{N}(\boldsymbol{r}_k - \boldsymbol{r}_{mean})^2 \right\rangle}$$

where $\boldsymbol{r}_k$ is the position of each $k^{th}$ heavy atom of the $PEG_{500}$ chain, $\boldsymbol{r}_{mean}$ is the position of the center of mass of the chain and $N$ is the total number of heavy atoms.

### 2.2.5 Self-diffusion coefficient

To evaluate the diffusional properties of cRGD molecules, either free to diffuse in water or conjugated to $PEG_{500}$ chains grafted on the $TiO_2$ NP surface, we estimated the self-diffusion coefficient based on the Einstein relation that relates the MSD to the self-diffusion coefficient $D$ as follows:

$$MSD = 2nDt$$

where *n* is the desired dimensionality and *D* is estimated by fitting the MSD at intervals in which a linear dependency with time *t* is observed.

**2.2.6 Polymer volume fraction**

Polymer volume fraction is defined as the fraction of the total volume occupied by the polymer. It was calculated using 0.1 Å wide spherical layers starting from the geometrical center of the NP. Each -$CH_2$- group and O atom was assigned a volume of 20 Å$^3$ and each water molecule a volume of 30 Å$^3$, as previously done in Refs. [45,59].

**2.2.7 Statistical analysis**

All the results reported in the tables are presented as mean ± standard deviation, while the error bars in the plots correspond to the standard deviation.

## 3. Results and Discussion

### 3.1 Building up the targeting nanodevice model

In the subsequent sub-sections, we will provide all the relevant details for each modeling step followed in the building-up process of the cRGD-conjugated PEGylated $TiO_2$ nanosystem model. Each CHARMM-based molecular component of the nanodevice will be subjected to preliminary MD simulation and posterior validation against available literature data. Fig. 1 shows the complete cRGD-conjugated PEGylated $TiO_2$ nanosystem model in the center (Section 3.1.6) and each of its components that will be investigated below in the clockwise order: cRGD ligands (Section 3.1.1), $PEG_{500}$ chains (Section 3.1.2), $PEG_{500}$-cRGD chains (Section 3.1.3), the spherical anatase $TiO_2$ nanoparticle (Section 3.1.4), and the same nanoparticle functionalized with 50 $PEG_{500}$ chains (Section 3.1.5).

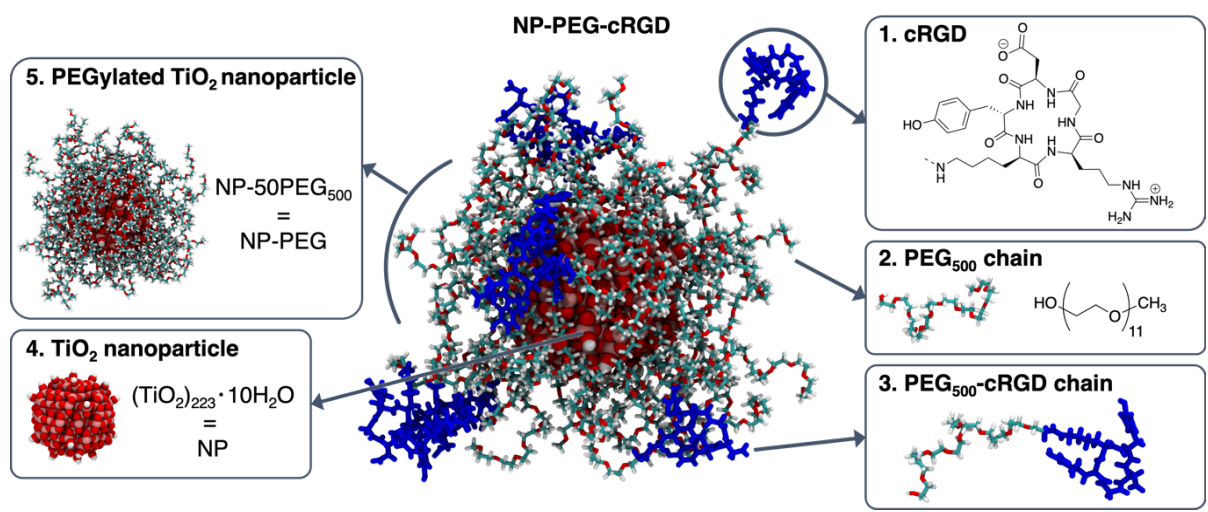

**Fig. 1.** cRGD-conjugated PEGylated TiO$_2$ nanosystem (at the center) and its components, in clockwise order, from top-right corner: 1. cRGD ligand, 2. PEG$_{500}$ chain, 3. PEG$_{500}$-cRGD chain, 4. TiO$_2$ nanoparticle and 5. PEGylated TiO$_2$ nanoparticle. Titanium, oxygen, carbon and hydrogen atoms are represented in pink, red, cyan and white, respectively. Spheres of vdW radius are used for TiO$_2$ NP atoms, licorice representation is used for PEG chains, and cRGD ligands are highlighted in blue.

### 3.1.1 cRGD model

In the following, we discuss the relevant chemical aspects to be considered for the accurate modeling of cRGD and we validate the CHARMM-based cRGD model through an extensive conformational and structural comparison against available experimental data [60]. In particular, we underline three major features of the cRGD molecule, which could be underestimated at a first sight:

1. The amino acid chirality: cRGD stands for c(RGDyK), i.e. cyclic Arg–Gly–Asp–D-Tyr–Lys, a cyclic peptide composed of four natural L-amino acids and one unnatural D-amino acid (D-Tyr) that increases the affinity towards integrins [61].

2. The protonation state of ionizable groups: the protonation state of cRGD at physiological pH conditions is shown in Fig. 2a according to the standard pKa values of amino acids in an aqueous solution. While arginine and lysine are protonated, aspartate is deprotonated, leading to an overall net charge of +1, which is neutralized by a chloride counterion to keep the systems' electroneutrality.

3. Peptide *cis-trans* conformations: both *cis* and *trans* are possible conformations for the amide bonds. However, the *trans* conformation is the most stable due to steric and electronic factors. Besides, according to the available literature, the *trans* conformation is the most stable one in proteins [62]. This is also confirmed by our simulated annealing calculations (see Supporting Material, Fig. S3). Even when the starting-point structures of cRGD contained Gly-Asp and Lys-Arg *cis* bonds, we observe that, after the annealing cycle, the most stable cRGD conformation is all-*trans*. Thus, the latter conformation is adopted throughout this work.

To assess the structural convergence and the most recurrent conformational states acquired by cRGD in water, we analyze the RMSD distribution for the whole cRGD structure and its amino acid side-chains (Fig. 2b). Fig. 2a illustrates cRGD structure, the definition for the Ψ (N-$C_\alpha$-C-N) and Φ (C-N-$C_\alpha$-C) torsional angles in cRGD, and the color code assigned for its side-chains (the same color codes are consistently adopted throughout the text). Fig. 2c shows the Ramachandran plots for the backbone torsional angles of cRGD analyzed over the MD production phase.

Upon analysis of Fig. 2b, we see that within the converged RMSD region (last 50 ns of the MD simulation), i.e. the production phase, the cyclic portion of cRGD (green) alternates between two main structural conformations and the most mobile sidechains are those of Arg and Lys.

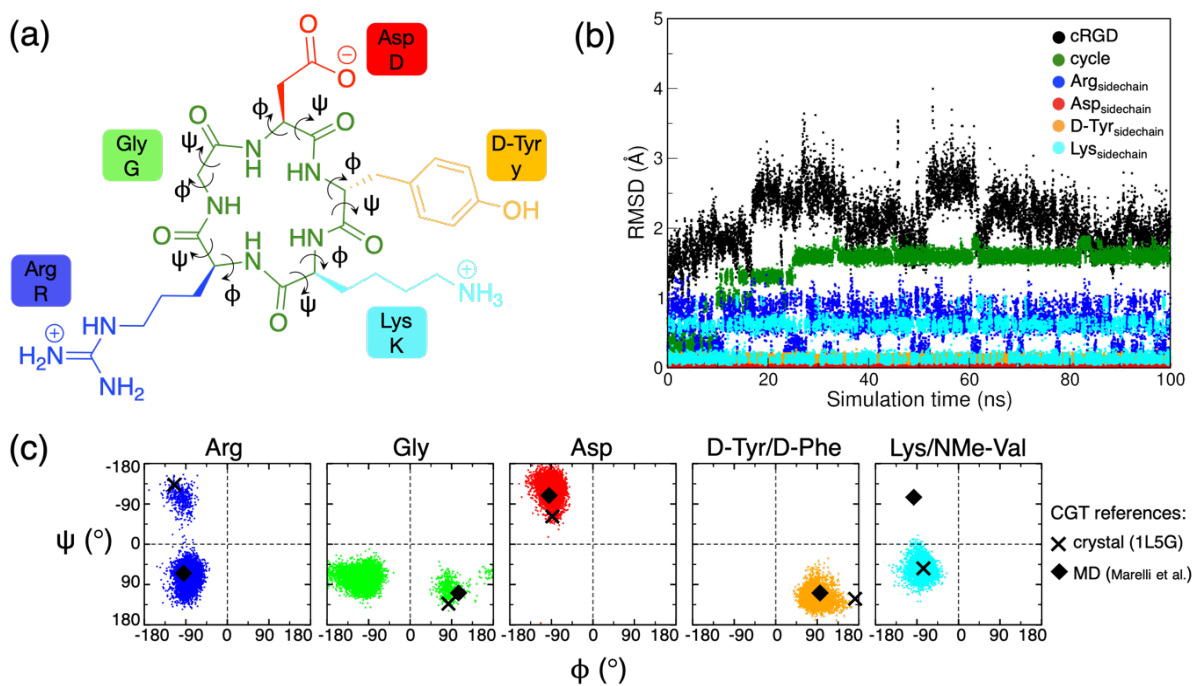

**Fig. 2.** (a) Structural formula of cRGD molecule with names and codes of its composing amino acids. Each amino acid is associated with the same color throughout the text. cRGD is shown in its protonation state at physiological pH conditions and colored accordingly to the RMSD plot and Ψ and Φ dihedrals definitions. (b) RMSD distribution of heavy atoms in the (i) whole cRGD structure, (ii) cRGD cycle, and (iii) each cRGD sidechain over the entire MD simulation. (c) Ramachandran plots for each amino acid residue of cRGD during the MD production phase (last 50 ns). Black crosses identify the dihedral angles of CGT found in the crystal structure of the complex $\alpha_v\beta_3$ integrin-CGT (PDB code: 1L5G [60]), while black diamonds are used for dihedral values found by Marelli et al. [63] from MD simulations of a free CGT in water. D-Phe and NMe-Val are CGT amino acids occupying cRGD D-Tyr and Lys place, respectively.

From the distribution of the torsional angles Ψ and Φ in the cRGD cycle in Fig. 2c, we find the presence of two main conformational states for the cRGD cyclic portion, mainly driven by the structural changes of Arg and Gly. Moreover, we compare our MD predictions with two sets of dihedral angle values reported in the literature for Cilengitide (CGT), which is a cyclic peptide (Arg–Gly–Asp–D-Phe–NMe-Val), whose chemical composition resembles the one of cRGD studied in this work and thus provides a fair reference. Black crosses and black diamonds refer to the crystal structure of CGT in complex with the $\alpha_v\beta_3$ integrin (PDB code: 1L5G [60]) and to MD predictions made by Marelli et al. [63] for CGT in water, respectively. The comparative analysis shows that the single cRGD ligand in the bulk-water phase is able to sample the backbone torsional dynamics in close agreement with the experimental crystal data. In addition, we also identify the sampling of a second dihedral angle distribution for Arg and Gly, which is not identified in the reference X-ray structure [60], but it is in the MD reference [63] in the case of Arg.

To identify the most populated conformations sampled during the MD production phase by the cRGD sidechains, we carry out the *cluster analysis* based on the RMSD of all heavy atoms, as detailed in Section 2.2.1. We find 8 clusters and report the cluster occupancy of the first three most populated clusters in Fig. 3a, since the other 5 clusters represent less than 1% of the conformations adopted during the production phase.

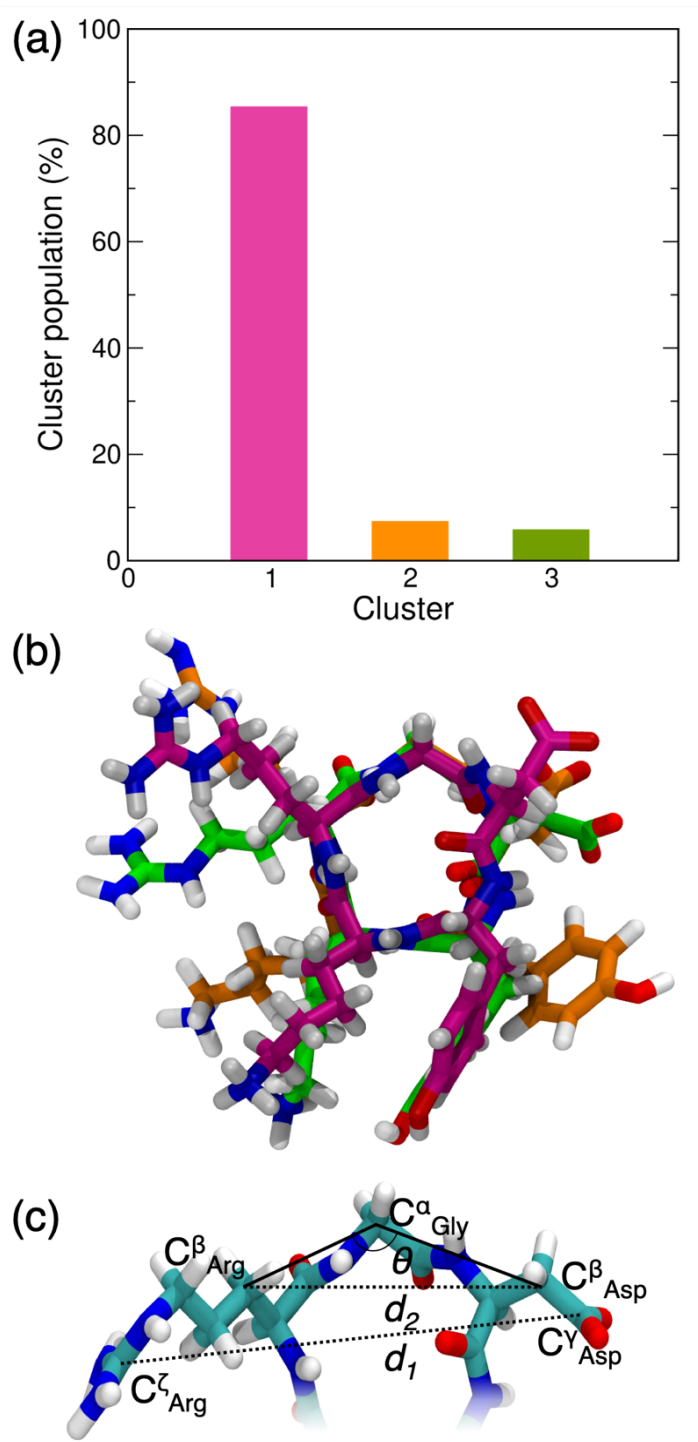

**Fig. 3.** (a) Cluster occupancy of the first three clusters found in the production phase of cRGD MD simulation in water using the GROMOS method, a cutoff of 2.0 Å and considering only heavy atoms. (b) Representative structure of cluster 1, cluster 2, and cluster 3, where C atoms are of the same color of the cluster, while oxygen, nitrogen, and hydrogen atoms are shown in red, blue, and white, respectively. (c) Definition of the three criteria used to assess cRGD potential to be biologically active[64,65].

Fig. 3b shows the representative structures of clusters 1, 2 and 3, i.e., the most recurrent conformations adopted by cRGD during the MD production phase. They share a similar conformation, especially in terms of the orientation of Arg and Asp sidechains. Only D-Tyr sidechain, which is not part of the RGD sequence, orients differently in the two cases.

Previous experimental and computational findings [64–67] have proven a direct link between the biological activity of RGD peptides and some of their intrinsic structural parameters. Three criteria were proposed for the assessment of biological activity of RGD-containing peptides [64,65] and they are illustrated in Fig. 3c: (i) the distance $d_1$ between the positively (Asp) and negatively (Arg) charged groups, i.e., between the $C^{\zeta}_{Arg}$ and the $C^{\gamma}_{Asp}$; (ii) the distance $d_2$ between the $C^{\beta}$ of Arg and Asp; (iii) the angle $\theta$ formed between the $C^{\beta}_{Arg}$ and the $C^{\beta}_{Asp}$ with the $C^{\alpha}_{Gly}$ as the vertex.

**Table 1.** The three criteria estimated from the CGT crystal structure [60], the optimized DFT structure of cRGD [67], and the three most populated clusters of cRGD structures sampled during the MD production phase.

| Structural parameter | CGT (Crystal) [60] | cRGD (DFT) [67] | Cluster 1 | Cluster 2 | Cluster 3 |
|---|---|---|---|---|---|
| $d_1$ (Å) | 13.7 | 13.4 | 11.4 (±1.4) | 10.8 (±1.7) | 12.3 (±1.6) |
| $d_2$ (Å) | 8.9 | 5.7 | 7.6 (±0.6) | 7.8 (±0.5) | 7.8 (±0.6) |
| $\theta$ (°) | 136 | - | 111 (±12) | 115 (±11) | 113 (±13) |

Overall, we find a fair agreement for $d_1$ and $d_2$ between our MD predictions (cRGD structure in Clusters 1, 2 and 3) and those found for CGT in the crystal structure [60] and by DFT calculations [67]. On the contrary, we find a considerable deviation between the cRGD structure in Clusters 1, 2 and 3 and the CGT structure in complex with $\alpha_v\beta_3$ integrin for $\theta$, as seen in Table 1.

These results above confirm the reliability of the CHARMM-based cRGD model to adequately sample the backbone conformational transitions and relevant geometrical parameters that may drive a successful binding to $\alpha_v\beta_3$ integrin. In Section 3.2.3, the discussion is resumed to investigate how the cRGD attachment impacts the above-cited geometrical criteria for a pre-evaluation of its bioactivity.

### 3.1.2 PEG$_{500}$ chain model

Herein, we will check the capability of the CHARMM-FF to accurately describe relevant experimental polymeric parameters of PEG$_{500}$ in an aqueous solution against previous predictions using the AMBER-FF [45,59], which were extensively validated against experimental data [68,69].

The polymeric chain model considered in this study contains 11 monomeric units with an overall molecular weight of 500 Da, thus PEG$_{500}$, and it is terminated with a hydroxyl group on the one side, which will be used to graft the chain to the NP surface, and a methyl group on the other side. To assess the correctness of the empirical FF parameters for the classical description of PEG$_{500}$ chains in water, we estimate two polymeric properties that are often quantified experimentally, namely end-to-end distance ($\langle h^2 \rangle^{1/2}$) and radius of gyration ($R_g$), comparing them against available theoretical and experimental literature data [45,59,68,69].

On the basis of the measurement of $\langle h^2 \rangle^{1/2}$ distance, the PEG$_{500}$ molecule described by the CHARMM-FF is, on average, slightly more extended (16.1 (±5.5) Å) in water than the one simulated using AMBER FF (14.3 (±4.0) Å) [59]. Moreover, we find a fair agreement for the $R_g$ between the CHARMM-based PEG$_{500}$ model and available numerical MD simulation [45] and experimental measurements [59,68,69] for an identical PEG$_{500}$ chain (Table 2).

**Table 2.** Radius of gyration for a single PEG$_{500}$ in water of a single PEG$_{500}$ chain in water over 50 ns of MD production phase and reference computational and experimental data with their relative standard deviation.

| Reference | $R_g$ (Å) |
|---|---|
| This work | 6.9 (±1.0) |
| MD [45] | 6.0 (±0.8) |
| Exp. [59][+] | 9 (±1) |
| Exp. [68][*] | 8.1 (±1.7) |
| Exp. [69][†] | 6.2 |

[+]Calculated from the relation $R_h = 0.64\ R_g$, where $R_h$ is the hydrodynamic radius.
[*]Calculated according to the relation $R_g = 0.0215\ M_w^{(0.583 \pm 0.031)}$.
[†]Calculated according to the relation $R_g = 0.0202\ M_w^{0.550}$.

In addition, we also validate the PEG$_{500}$ model in the bulk-water phase for the CHARMM-FF through a comparative analysis of the *polymer dihedral distribution* against the same MD descriptor estimated using the AMBER-FF. We observe two maximum peaks for the O-CH$_2$-CH$_2$-O dihedral angle: one positioned at -69.9 ° and a second one at +72.1 ° (see Fig. S4 in Supporting Material). These estimations using the CHARMM-FF agree with those reported in Ref. [45] for the same system and in Ref. [70] and [71].

Finally, by analyzing the H-bonds established between the CHARMM-based PEG$_{500}$ chain and water molecules, we confirm that the chemical behavior of the ether groups in the chain in water is correctly described, with their oxygen atoms behaving as H-bond acceptors and water molecules as H-bond donors. We find an average number of H-bonds per monomeric unit of 0.81 (±0.3), which is close enough to that obtained using the AMBER-based PEG$_{500}$ model, i.e. 0.99 (±0.1) per monomeric unit.

Based on the results above, one can infer that CHARMM-FF is a reasonable FF choice to describe polymeric linkers based on polyethylene glycol units under explicit water solvation. In the next section, we will proceed with the validation of a PEG$_{500}$ molecule conjugated with cRGD.

### 3.1.3 PEG$_{500}$-cRGD chain model

After implementing and validating the cRGD and PEG$_{500}$ components separately in Sections 3.1.1 and 3.1.2, we now conjugate both cRGD and PEG$_{500}$ models into a single molecule through the formation of an amide bond, as seen in Fig. 4a, mimicking the experimental structure found in [25]. The resulting zwitterionic cRGD-conjugated PEG$_{500}$ chain model (PEG$_{500}$-cRGD) will be used here to functionalize the TiO$_2$ NP following the protocol reported in Section 2.1.2.

To find out whether the conjugation of cRGD to the PEG$_{500}$ chain affects the ligand conformation or not, we repeat the same conformational analysis as previously done for a single cRGD in Section 3.1.1. Fig. 4b displays the Ramachandran plot analysis for Φ and Ψ torsional angles of a cRGD ligand conjugated to the PEG$_{500}$ chain.

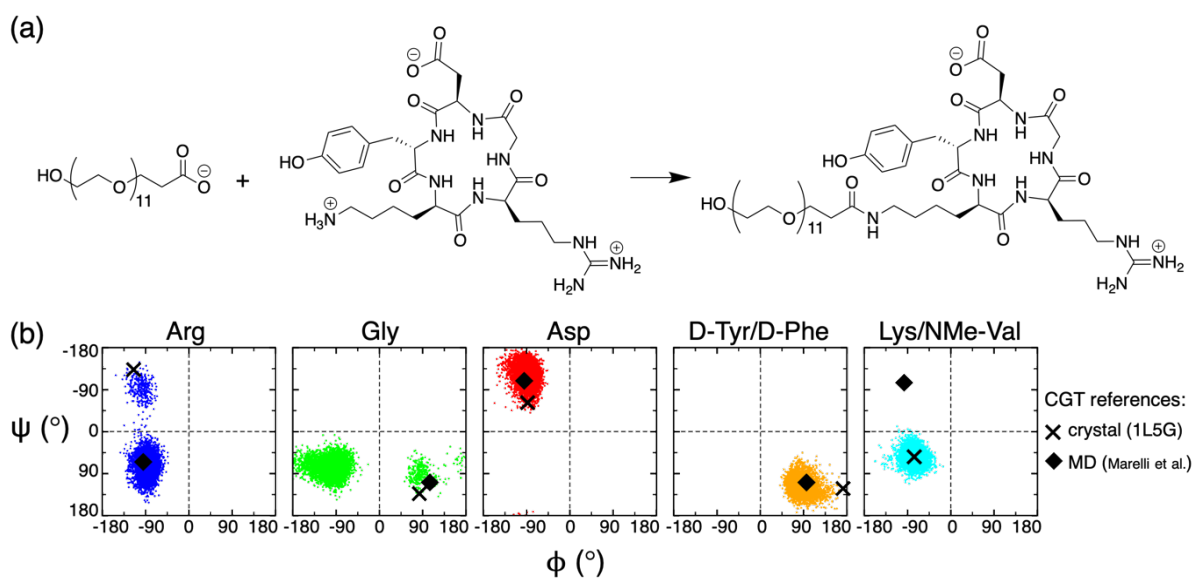

**Fig. 4.** (a) Conjugation of a cRGD ligand to a PEG$_{500}$ chain resulting in a cRGD-conjugated PEG$_{500}$ chain. (b) Ramachandran plots for each amino acid of cRGD in the cRGD-conjugated PEG$_{500}$ chain during the MD production phase. Black crosses identify the dihedral angles of CGT found in the crystal structure of the complex α$_v$β$_3$ integrin-CGT (PDB code: 1L5G [60]), while black diamonds refer to the dihedral values found by Marelli et al. [63] from MD simulations of a free CGT in water. D-Phe and NMe-Val are CGT amino acids occupying cRGD D-Tyr and Lys place, respectively.

Upon analysis of the Ramachandran plot in Fig. 4b, we can observe a resemblance between the conformational states acquired by the cRGD ligand, whether conjugated to PEG$_{500}$ chains or not (Fig. 4b vs. Fig. 2c). Thus, we can infer that cRGD conjugation to PEG$_{500}$ linkers has no impact on the ligand's torsional dynamics. Overall, we confirm the reliability of the CHARMM-FF parameters to describe the torsional dynamics of cRGD conjugated to PEG$_{500}$ chains in explicit bulk-water solvation and their suitability to compose more complex systems, as will be done in Section 3.1.6.

### 3.1.4 TiO$_2$ nanoparticle model

The spherical TiO$_2$ NP model (diameter 2.2nm, (TiO$_2$)$_{223}$·10H$_2$O) used in this study has been developed by some of us in previous works using high-level quantum mechanical (QM) methods [46,47]. This model was proven to be chemically stable upon water adsorption [72] and solvation using QM and QM/MM approaches [73].

Moreover, the validity and accuracy of the chosen force field parameters [50] for the description of bioinorganic nanohybrids was also assessed in previous works by some of us using QM and QM/MM calculations [51].

### 3.1.5 PEGylated TiO$_2$ nanoparticle model

Before approaching the final cRGD-conjugated PEGylated TiO$_2$ nanosystem, we first validate its non-conjugated counterpart in water, namely the PEGylated TiO$_2$ NP with 50 PEG$_{500}$ chains (or NP-PEG). The main motivations are the following: (i) access the reliability of the chosen FF parameters for the PEGylated TiO$_2$ NP model to reproduce available reference from theoretical and experimental data in the literature [45,59]; (ii) use this NP model as the substrate to build up the final cRGD-conjugated nanodevice; (iii) adopt this model as the reference model to further investigate the effects of ligand density.

Several experimental studies have proven that moderate PEGylation (PEG chains of $\overline{MW} = 2k\,Da$) shows the best compromise between an anti-opsonization strategy and favorable cell uptake [74,75], although it is still controversial in the experimental literature [76,77]. However, the full-atomistic modeling and simulation of such large systems would become impractical to achieve satisfactory sampling and statistics. Hence, we decided to use short PEG chains ($\overline{MW} = 500\,Da$), which are believed to prevent ligand entangling and cloaking [10,77], and a high grafting density regime on a relatively small TiO$_2$ NP (diameter of 2.2 nm).

The choice of the NP size is also a crucial parameter. It is noteworthy that, when it comes to a curved surface (e.g., highly curved nanoparticles) as the substrate for PEGylation, experimental findings suggest that the achievement of brush-like conformational regimes is not solely dependent on the specific molecular weight of PEG chains but also on the nanoparticle size [78], extent of polymeric packing (PEG-PEG spacing) [79], and grafting density of PEG chains tethered on the NP surface [59].

Therefore, we validate the CHARMM-based PEGylated TiO$_2$ NP model against available AMBER-based simulations by some of us [45] as previously done in Section 3.1.2, i.e., by quantifying $\langle h^2 \rangle^{1/2}$ distance and $R_g$. Besides, we also validate our PEGylated TiO$_2$ NP model for the Mean Distance from the Surface (*MDFS*) and number of H-bonds descriptors, whose definitions can be found elsewhere [45].

In Table 3, we compile the structural parameter data estimated using the CHARMM-based PEGylated TiO$_2$ NP model and the available reference data. For all the polymeric parameters analyzed here, we observe a fair agreement between our MD results and the simulation predictions in the literature.

**Table 3.** Structural polymeric parameters of NP-PEG during the last 50 ns of MD production with their standard deviation.

| Structural Parameter | Grafting density (chain Å$^{-2}$) | $\langle h^2 \rangle^{1/2}$ (Å) | $R_g$ (Å) | MDFS (Å) | N. of H-bonds |
|---|---|---|---|---|---|
| This work | 0.02252 | 17.0 (±0.5) | 6.7 (±0.1) | 11.6 (±0.2) | 0.75 (±0.3) |
| MD [45] | 0.02252 | 18.2 (±0.4) | 6.8 (±0.1) | 11.1 (±0.1) | 0.56 (±0.2) |

To characterize the polymer conformation of PEG chains at the grafting density regime ($\sigma = 0.02252\ chain\ \text{Å}^{-2}$) adopted in the PEGylated TiO$_2$ model studied here, we compare our MD results against the theoretical predictions by the Daoud and Cotton model [80] for polymer-grafted nanoparticles (Fig. 5). This analytical model predicts that star-like polymers at a high grafting density present a central rigid core with constant polymer density, surrounded by a concentrated regime where the polymer volume fraction $\Phi(r)$ depends linearly on the radial distance from the NP center ($r^{-1}$) (Fig. 5, red dashed line), followed by a semi-diluted polymer brush regime in which $\Phi$ behaves as $\Phi \propto r^{-4/3}$ (Fig. 5, blue dashed line).

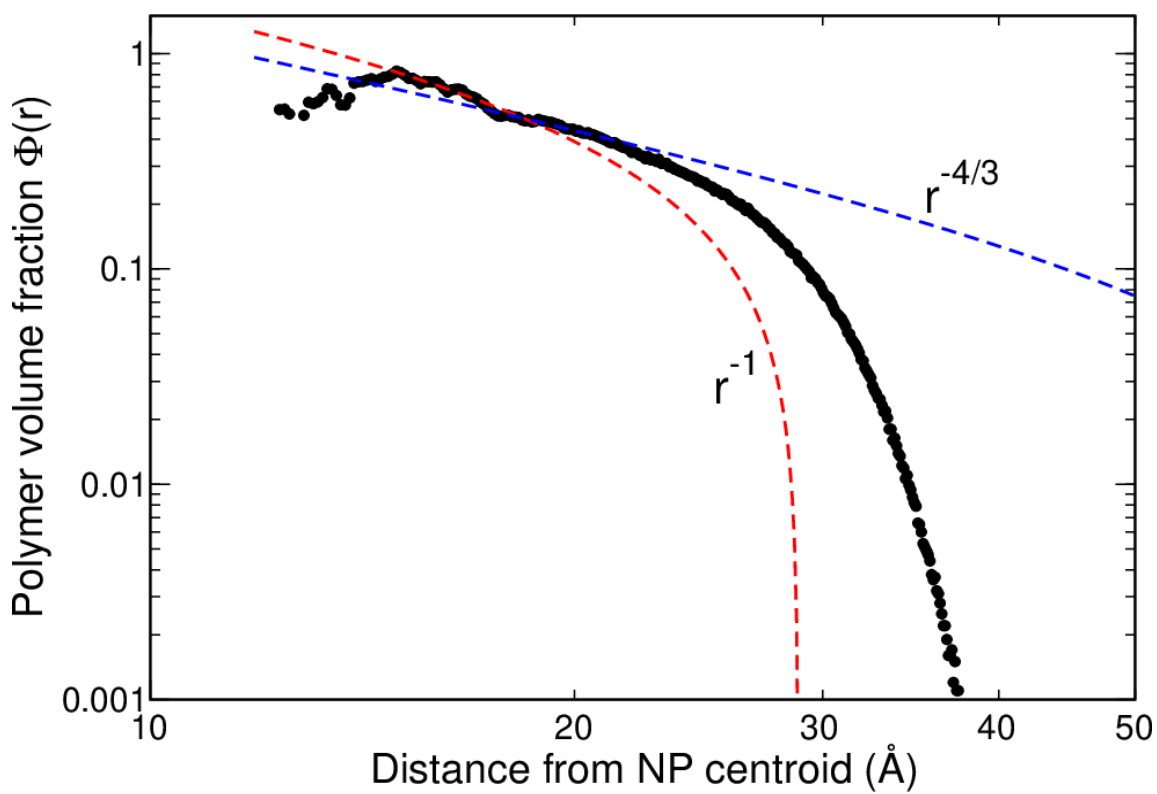

**Fig. 5** Log-log plot of MD predictions for the polymer volume fraction of PEG chains tethering the reference PEGylated 2.2 nm TiO$_2$ nanoparticle from the NP geometric center towards the bulk-water phase. The red and blue dashed lines correspond to the Daoud-Cotton model predictions for Φ(r) in the concentrated (Φ ∝ $r^{-1}$) and semi-diluted (Φ ∝ $r^{-4/3}$) brush regime, respectively.

Upon analysis of Fig. 5, we observe that both the concentrated and the semi-diluted brush regimes, theoretically predicted by the Daoud and Cotton model, are respected by the MD predictions for the reference PEGylated TiO$_2$ model in the ranges of 15.0-17.5 Å and 17.5-22.0 Å, respectively. This observation confirms that low-weight PEG chains tethered on a small, highly curved TiO$_2$ NP at high grafting density may induce brush conformations up to a few nanometers from the NP surface. However, special care should be taken when considering similar PEGylated models for protein adsorption studies, which might require longer PEG chains to broaden the brush-like PEG region.

### 3.1.6 cRGD-conjugated PEGylated TiO$_2$ nanoparticle model: the complete nanodevice

At this point, we are finally in the position to assemble all the single components and their molecular combinations to build up the complete nanodevice. We uniformly distribute the cRGD ligands over the PEGylated NP surface, keeping them as far as possible from each other. To that end, we conjugate a suitable number of cRGD ligands with the terminal group of $PEG_{500}$ chains (that are attached to the $TiO_2$ NP surface). The rationale behind the chosen cRGD densities relies on previous experimental attempts to predict the optimal density of cRGD to be conjugated with the PEGylated NP for enhanced on-targeting of integrins $\alpha_v\beta_3$ in tumor cells [8,10,11]. Table 4 reports the number of cRGD ligands per nanoparticle, the degree of cRGD ligand conjugation expressed as the percentage of conjugated PEG chains out of the total number of PEG chains × 100 (this % is inserted in the label NP-PEG-cRGD% used throughout the text to identify the three models under investigation), and the cRGD ligand density expressed as the number of cRGD ligands per unit of the nanoparticle surface area. We choose $nm^2$ as the unit of surface area to allow a direct and straightforward comparison with the experimental ligand densities reported in Refs. [7,10,11,81,82]. The surface area of the 2.2 nm $TiO_2$ nanoparticle model is taken from Ref. [47]. Three different cRGD densities are modeled: 1) 10% of cRGD-conjugated $PEG_{500}$ chains corresponding to 0.2 ligand $nm^{-2}$ (from now on NP-PEG-cRGD10 model in Fig. 6a); 2) 20% of cRGD-conjugated $PEG_{500}$ chains corresponding to 0.5 ligand $nm^{-2}$ (from now on NP-PEG-cRGD20 model in Fig. 6b); 3) 50% of cRGD-conjugated $PEG_{500}$ chains corresponding to 1.1 ligand $nm^{-2}$ (from now on NP-PEG-cRGD50 model in Fig. 6c).

**Table 4.** Correspondence between the number of cRGD ligands per nanoparticle, the degree of ligand conjugation expressed as a percentage, and the ligand density expressed as the number of cRGD ligands per unit of nanoparticle surface area.

| Model | NP-PEG-cRGD10 | NP-PEG-cRGD20 | NP-PEG-cRGD50 |
|---|---|---|---|
| **Number of cRGD per NP** | 5 | 10 | 25 |
| **Degree of cRGD conjugation (%)**[*] | 10 | 20 | 50 |
| **Ligand density (ligand $nm^{-2}$)**[**] | 0.2 | 0.5 | 1.1 |

*Degree of cRGD conjugation refers to the percentage of conjugated cRGD-PEG chains over all grafted chains on the NP surface.

**Ligand density refers to the number of cRGD ligands per unit of nanoparticle surface area.

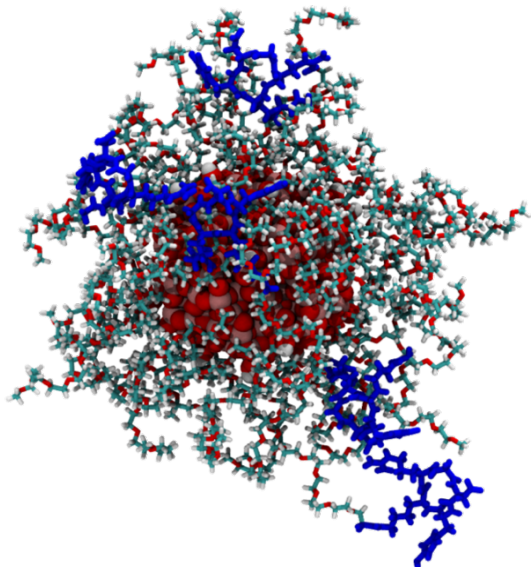

(a) NP-PEG-cRGD10

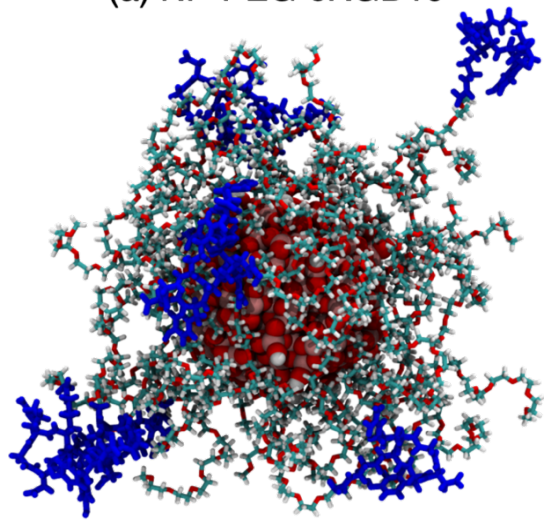

(b) NP-PEG-cRGD20

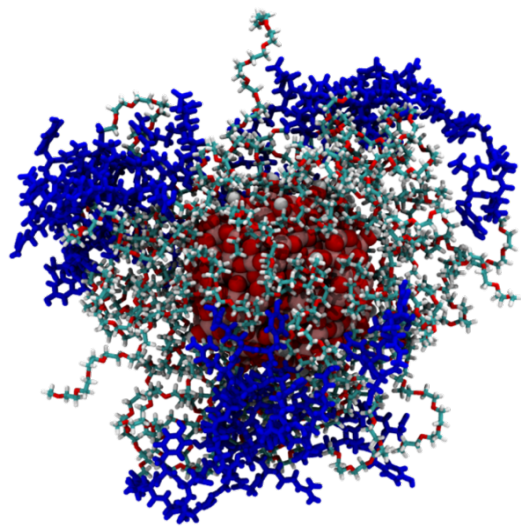

(c) NP-PEG-cRGD50

**Fig. 6.** Final snapshot of the MD simulation of (a) NP-PEG-CRGD10, (b) NP-PEG-CRGD20 and (c) NP-PEG-cRGD50 models. Water is not represented for the sake of clarity. Titanium, oxygen, carbon and hydrogen atoms are represented in pink, red, cyan and white, respectively. Spheres of vdW radius are used for TiO$_2$ NP atoms, licorice representation is used for PEG chains, and cRGD ligands are highlighted in blue.

## *3.2 Analyzing the nanodevice: impact of cRGD ligands density*

In the following subsections, we investigate the role of ligand density on some biophysical properties of relevance to the cRGD ability to target the α$_v$β$_3$ integrin. In Section 3.2.1, we explore the role of ligand density on their conformational state and extent of exposure (also referred to as "RGD presentation"). Next, in Section 3.2.2, we analyze the interplay between the ligand density and the stability of cRGD-conjugated nanodevices in the bulk-water phase. Also, the effect of covalently fettering the cRGD ligand at the end of PEG$_{500}$ chains on the ligand's diffusional properties is brought to attention. Finally, in Section 3.2.3, we shed light on the possible implications of cRGD conformational changes for successful integrin binding.

### 3.2.1 On the cRGD presentation

Cell adhesion is of utmost importance to improve the target-specific delivery of cRGD-conjugated nanomaterials, which directly depends on the degree of targeting-ligands exposure and their orientation out of the "stealth" polymeric coating layer towards the biological milieu. Experimental findings have suggested that the spatial organization of ligands plays a crucial role in the T-cell immune response [83]. The authors discovered that ligand's presentation and disposition, at optimal nanometric distances from each other on DNA flat sheets, have the ability to suppress T-cell signaling *in vivo* [83]. Moreover, a recent combined experimental and MD simulation study has shown the implications of cRGD presentation and orientation on the targeting efficiency of cRGD-conjugated PEGylated Au NPs towards α$_v$β$_3$ integrin. They found that the enlargement of ligand-conjugated chains by a short molecular spacer is sufficient to enhance the cRGD moiety presentation out of the polymeric stealth layer, which is corroborated by a substantial increase in the water accessibility towards the cRGD surface [13].

Here, we investigate how different degrees of ligand conjugation impact the cRGD presentation and modulate the ligand's molecular environment in the system. To quantify the extent of cRGD exposure from the stealth PEG region towards the bulk-water phase, we first compute the number density profiles for the relevant moieties of the NP-PEG-cRGD10, NP-PEG-cRGD20 and NP-PEG-cRGD50 nanosystems. Fig. 7 shows the number density profiles for water molecules, PEG$_{500}$ chains and the C.O.M. (center of mass) of cRGD ligands from the geometric center of the nanodevice, computed with a binning size of 0.5 Å.

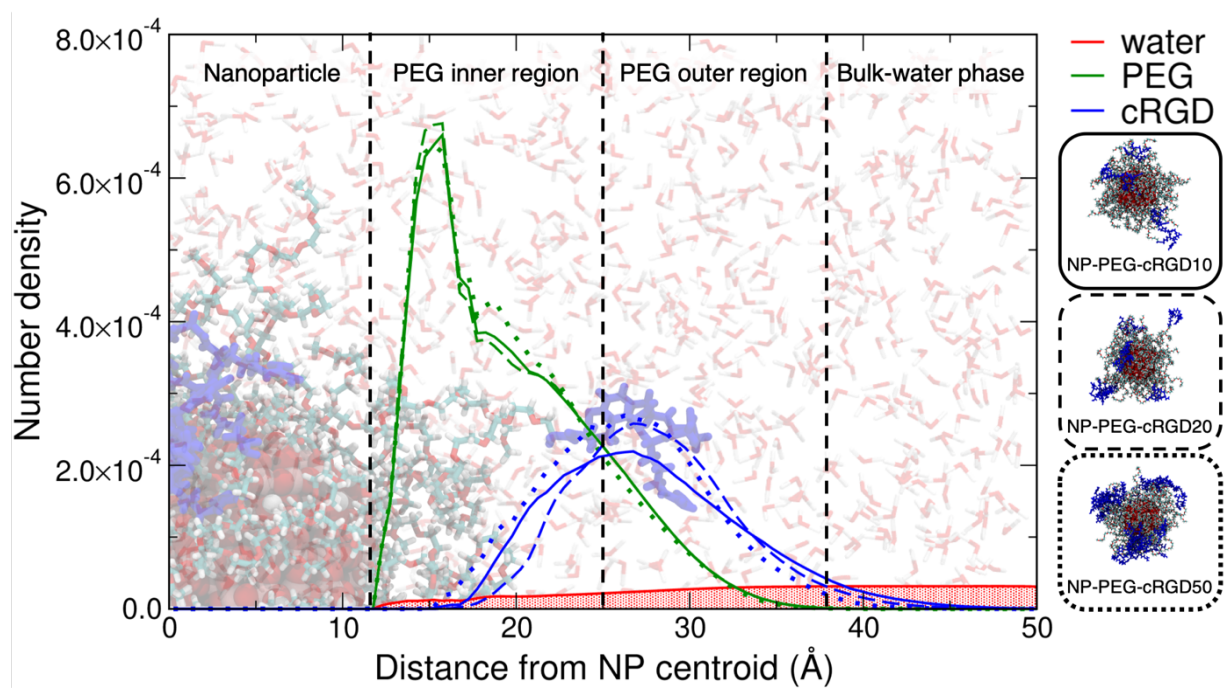

**Fig. 7.** Number density of water (red), PEG$_{500}$ chains (green), and the C.O.M. of cRGD ligands (blue) from the geometric center of the NP. Solid lines refer to NP-PEG-cRGD10 system, dashed lines to NP-PEG-cRGD20 system, and dotted lines to NP-PEG-cRGD50 system. The background shows the TiO$_2$ nanoparticle surrounded by PEG chains, eventually conjugated to cRGD, and immersed in water. A binning size of 0.5 Å is used.

Number density profiles in Fig. 7 show that, independently from the ligand density, cRGD ligands are mostly located at the interface between the PEGylated NP surface and the aqueous phase. This observation remarkably agrees with what was found experimentally by Valencia et al. [11], who observed that RGD ligands are predominantly exposed on the surface of PEGylated NPs over a wide range of ligand densities. Interestingly, we identify a direct correlation between the degree of PEG$_{500}$ chains conjugated with cRGD ligands and the distribution of cRGD ligands in the system (Fig. 7). At 10% of cRGD conjugation, the cRGD density has a log-normal distribution with the longest tail towards the bulk-water phase among all systems. The latter observation implies that cRGD ligands can reach farther regions beyond the PEG$_{500}$ layer with considerable exposure to the bulk-water phase. By increasing the cRGD conjugation density from 10% to 20%, we identify an increased cRGD density in the outer region of the PEG$_{500}$ shell, confirmed by a slight shift in the density maximum peak towards the bulk-water phase in the latter system. Further raising of the cRGD density from 20% to 50% shows no improvement in the exposure of cRGD towards the bulk-water phase. On the contrary, we see an increasing number of cRGD ligands located in the inner region of the PEG$_{500}$ layer, with the resulting number density profile mostly resembling that estimated at 20% of cRGD density.

Although the number density profiles can accurately predict the average location of cRGD in the system, we still have no information on the orientation and entangling tendencies of cRGD at different ligand densities. To this end, we analyze the tilt angle acquired by cRGD against their distance from the NP center (Fig. 8). For the angular analysis, two vectors are considered: one vector is defined from the NP center to the C$^{\alpha}_{Lys}$ and the second vector from the C$^{\alpha}_{Lys}$ (vertex) to the N atom of Asp in the cRGD cycle (Fig. 8a).

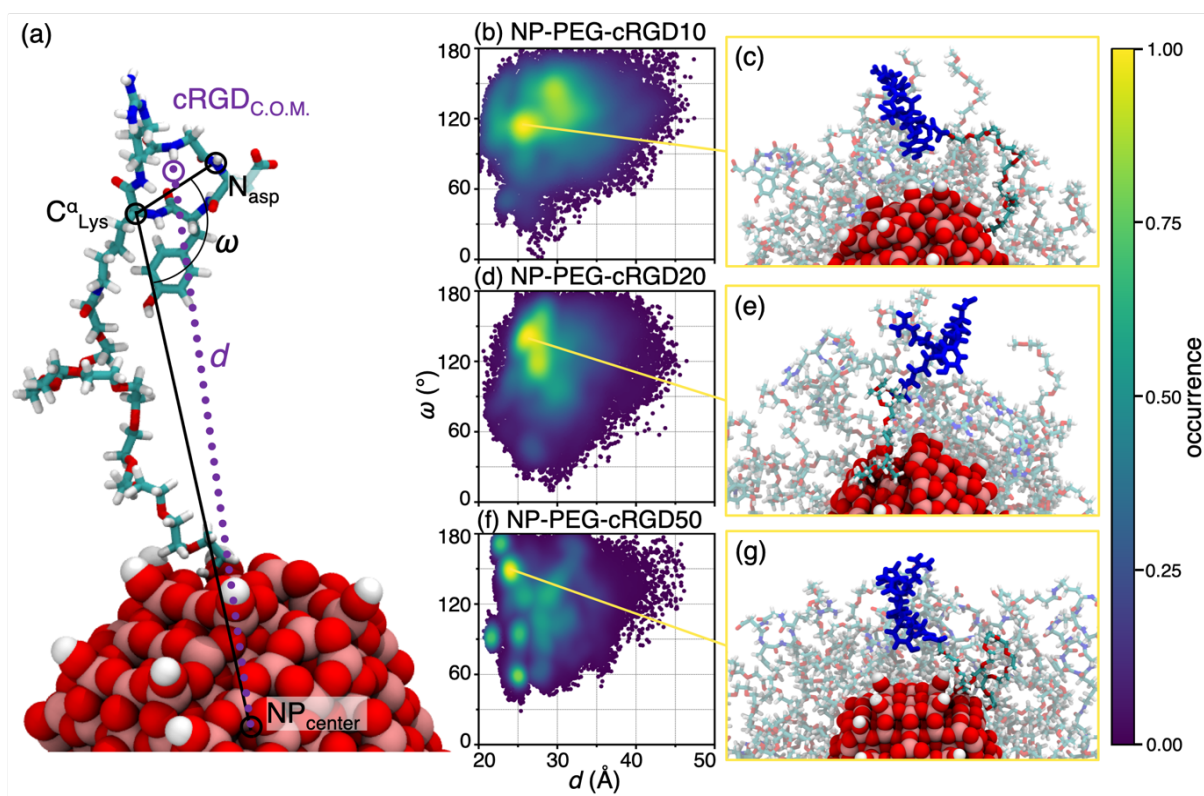

**Fig. 8.** Contour plot of angle vs. distance distributions over the last 50 ns of MD production. (a) Definition of the angle $\omega$ (NP$_{center}$-C$^{\alpha}_{Lys}$-N$_{Asp}$, black) and the distance $d$ (NP$_{center}$-cRGD$_{C.O.M.}$, purple) considered. (b) Contour plot for NP-PEG-cRGD10 nanodevice with one of the most recurrent orientations for cRGD (c). (d) Contour plot for NP-PEG-cRGD20 nanodevice with one of the most recurrent orientations for cRGD (e). (f) Contour plot for NP-PEG-cRGD50 nanodevice with one of the most recurrent orientations for cRGD (g). Titanium, oxygen, nitrogen, carbon, and hydrogen atoms are shown in pink, red, blue, cyan and white, respectively. TiO$_2$ NP atoms are represented with spheres of vdW radius, licorice representation is used for PEG and PEG-cRGD chains, and cRGD ligands are highlighted in blue. cRGD$_{C.O.M.}$ refers to the center of mass of heavy atoms of cRGD cycle.

Fig. 8 reveals that the ligand density modulation affects the degree of cRGD exposure and orientation. At 10% of cRGD conjugation, we find that stand-up configurations (tilt angles higher than 120°) are denser populated at farther distances from the NP surface (Fig. 8b) compared to the systems at higher ligand densities (Fig. 8d and f), while at 20% of cRGD conjugation, we identify a clear tendency of cRGD in populating a straighter stand-up configuration at shorter distances, as seen in Fig. 8d. Further increase in the cRGD density leads to a substantial modification in the interplay between the cRGD molecules' orientation and their distance from the NP surface. Fig. 8f reveals multiple hotspots at closed-up configurations (tilt angles lower than 120°) being densely populated at 50% of cRGD conjugation, which does not occur at lower cRGD densities (Fig. 8b and 8d).

To unveil the molecular reasons behind this cRGD behavior noticed at different densities, we conduct a volumetric analysis to quantify the surface area of contact between cRGD and other nanosystem components (PEG$_{500}$ chains, solvent molecules, and other cRGD ligands) with the aid of Voronoi tessellation [84]. In this way, we can quantify the surface area exposure of ligands to distinct parts of the system and find their preferential molecular neighborhood along the MD trajectory. Table 5 shows the surface area of contact between the cRGD ligands at 10%, 20% and 50% of cRGD conjugation normalized by the total number of ligands.

**Table 5.** Mean surface area of contact (Å$^2$) of cRGD in the solvated systems and its standard deviation. The values are averaged over the MD production phase and normalized by the total number of cRGD ligands.

| Area of contact (Å$^2$) | NP-PEG-cRGD10 | NP-PEG-cRGD20 | NP-PEG-cRGD50 |
|---|---|---|---|
| cRGD – water | 578 (±38) | 544 (±25) | 440 (±9) |
| cRGD – PEG$_{500}$ | 99 (±30) | 119 (±24) | 112 (±8) |
| cRGD – cRGD | 6 (±8) | 13 (±5) | 136 (±6) |

Upon analysis of Table 5, we see that the molecular surface of cRGD ligands is mainly accessed by water molecules, although the solvent surrounding the ligand surface decreases with the increase of cRGD conjugation. The largest decrease of water contact with the cRGD surface occurs going from 20% to 50% of cRGD conjugation, corresponding to a reduction of about 100 Å$^2$ in the water accessibility towards the ligand surface. We may notice that, in opposite trend but as a compensating effect, the surface area of contact between cRGD ligands themselves gets larger to a similar extent (of about 120 Å$^2$). The reduction of cRGD ligand's solvation and the simultaneous increment of cRGD – cRGD surface area of contact is likely the primary mechanism behind the slight decrease of cRGD molecules on the surface of PEGylated NPs at high ligand regime as evidenced experimentally [11].

At first glance, the volumetric data (Table 5) seem to contradict the number density data reported in Fig. 7, where a higher cRGD density leads to an enhanced number of cRGD ligands populating the outer region of PEG$_{500}$. However, further volumetric analysis reveals a substantial enlargement in the surface area of contact between cRGD-cRGD ligands at higher cRGD densities conjugation (Table 5), reducing the solvent accessibility to their surface. Hence, the observation that cRGD ligands populate a more stand-up conformation at higher ligand density is likely linked to undesirable ligand-ligand interactions than their exposure to the bulk-water phase, which is rather evident when the cRGD density is the highest among all.

From the above results, we can infer that cRGD ligands are mainly located at the PEG$_{500}$-water interface, and their exposure to the bulk-water phase in an upward mode is preferred at the lowest ligand density. In addition, we also notice that the cRGD density directly impacts the ligand-ligand interactions, which may impair the water accessibility towards the ligand molecular surface.

### 3.2.2 On the nanodevice stability and diffusion in water

Recent experimental investigations have hypothesized a direct link between the targeting ligands solubility and their availability on the nanosystem surface. A recent experimental investigation has verified that targeting ligands with high water solubility tend to escape the hydrophobic polymeric region of PEGylated nanoparticles [11]. In the same vein, experimental findings have speculated that enhanced water solubility of polymeric chains may increase the RGD ligand diffusion and, therefore, impair the binding affinity of RGD to integrins [61,85]. A recent experimental work has also unveiled the role of ligand diffusion on specific $α_5β_1$ and $α_vβ_3$ integrin activation and signaling. These findings suggested an enhanced cell adhesion and differentiation by $α_vβ_3$ integrin on covalently-immobilized cRGD ligands compared to higher ligand diffusion regimes [86].

The interplay between the nanosystem's solubility and the ligand diffusion may have significant implications on the ability of cRGD to bind integrins, and therefore, we investigate the nanodevices' stability and the ligand diffusion upon increasing of cRGD density on its surface. To do so, we first estimate the intermolecular interaction energy between the NP-PEG-cRGD10, NP-PEG-cRGD20 and NP-PEG-cRGD50 nanosystems and water molecules. After that, we decompose the total energy of interaction into its respective electrostatic and vdW components, as detailed in Section 2.2.3.

Table 6 compiles the nonbonded interaction energy and its vdW and electrostatic contributions between each system and water and between different nanodevice components for the NP-PEG-cRGD10, NP-PEG-cRGD20 and NP-PEG-cRGD50 systems.

**Table 6.** Non-bonded interaction energy between NP-PEG-cRGD10, NP-PEG-cRGD20 and NP-PEG-cRGD50 nanosystems and water, cRGD ligands and water, cRGD ligands and PEG$_{500}$ chains, and ligands themselves. For cRGD interactions the data are normalized by the total number of cRGD ligands in the system. The average values over MD production phase and their standard deviations are reported.

| System | Interaction | Energy (kcal mol$^{-1}$) | |
|---|---|---|---|
| | | Coul. | vdW |
| **NP-PEG** | System – water | -4.61·10$^3$ (±8·10) | -1.62·10$^3$ (±4·10) |
| **NP-PEG-cRGD10** | System – water | -6.0·10$^3$ (±1·10$^2$) | -1.69·10$^3$ (±5·10) |
| | cRGD – water | -236 (±12) | -19 (±4) |
| | cRGD – PEG$_{500}$ | -4 (±4) | -9 (±3) |
| | cRGD – cRGD | -3 (±5) | -0.5 (±0.9) |
| **NP-PEG-cRGD20** | System – water | -7.1·10$^3$ (±1·10$^2$) | -1.71·10$^3$ (±5·10) |
| | cRGD – water | -226 (±8) | -17 (±3) |
| | cRGD – PEG$_{500}$ | -6 (±3) | -9 (±2) |
| | cRGD – cRGD | -6 (±3) | -1.3 (±0.7) |
| **NP-PEG-cRGD50** | System – water | -9.4·10$^3$ (±2·10$^2$) | -1.87·10$^3$ (±6·10) |
| | cRGD – water | -190 (±5) | -14 (±1) |
| | cRGD – PEG$_{500}$ | -2 (±1) | -5 (±1) |
| | cRGD – cRGD | -26 (±2) | -8.3 (±0.6) |

Analysis of Table 6 reveals that the main energetic stabilization comes from intermolecular interactions between the nanosystem and solvent. Interestingly, we identify that increasing the cRGD content considerably intensifies the energetic stabilization in water compared to the reference model (NP-PEG), being the electrostatic contribution the most relevant to it. On the other hand, the role of vdW interactions seems to be limited to a stabilization of the non-electrostatic forces between the nanosystem components, particularly between cRGD and PEG$_{500}$ chains. Noteworthy, the decreasing intensity of the electrostatic interactions between cRGD and water in Table 6 well agrees with the volumetric analysis in the previous section (Table 5), where we noticed a reduction of the surface area of contact between cRGD and the solvent molecules at increasing cRGD ligand density.

As seen in Table 6, the energetic stabilization coming from intermolecular interactions between the nanodevice's components are way less intense than those with the solvent. However, we identify that ligand-ligand interactions become relevant at 50% of cRGD conjugation mainly strengthened by electrostatic interactions, while they remain negligible at lower ligand density. Further discussion on the implications of ligand-ligand interactions on the optimal ligand presentation for integrin binding can be found in the previous section.

Finally, we investigate the effect of anchoring the cRGD ligands to the polymeric chains on their diffusional properties, which we expect to be reduced. To this end, we calculate the diffusion coefficient of cRGD either free to diffuse in bulk-water phase (reference) or bonded to a PEG$_{500}$ chain in the NP-PEG-cRGD10, NP-PEG-cRGD20, and NP-PEG-cRGD50 nanosystems (Fig. 9). The self-diffusion coefficient is estimated at MD trajectory intervals showing a linear dependence between MSD and time (for details see Section 2.2.5). The mean values and standard deviations can be found in Supporting Material in Table S3.

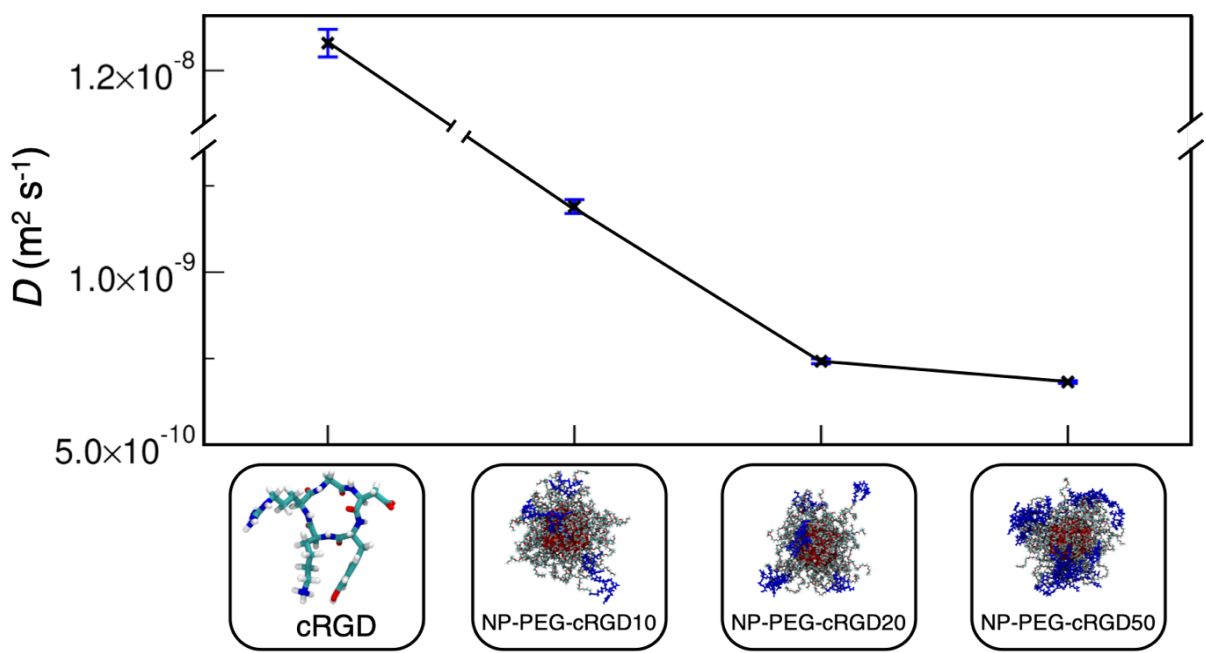

**Fig. 9.** Self-diffusion coefficients ($D$) of single cRGD in water and bonded to PEGylated $TiO_2$ nanodevices at 10%, 20% and 50% of cRGD conjugation with their error bars.

Comparing the self-diffusion coefficients in Fig. 9, we identify a decrease of one order of magnitude in the self-diffusion coefficient of cRGD molecules due to their conjugation to $PEG_{500}$ chains in the NP-PEG-cRGD10 system. Increasing the cRGD conjugation to 20% and 50% further slows down the cRGD self-diffusion coefficients, reaching values about two orders of magnitude smaller than those observed for the free cRGD molecule diffusing in water. This latter remark is intimately associated with the increasing of the cRGD-cRGD interactions identified in the previous volumetric and energetic analyses (Tables 5 and 6). Hence, we can infer that not only the ligand restraints due to their conjugation to polymeric chains but also the ligand density parameter affects the cRGD diffusion.

Based on the above results, we can infer that increasing the cRGD molecules' density on the nanosystem surface makes them more stable in water than their non-conjugated counterpart. Interestingly, our simulations establish that increasing cRGD conjugation extensively slows down the ligand self-diffusion until it reaches a critical ligand density of 0.5 ligand $nm^{-2}$. Above this limit, no significant impact on the ligand's self-diffusion has been observed. Recent experimental evidence [85] indicates that reduced ligand mobility enhances cRGD binding to integrins. Hence, we argue that moderate cRGD density might be sufficient to assure an improved targeting and avoid undesirable effects (e.g., ligand over-clustering) that might occur at higher ligand densities.

### 3.2.3 On the optimal conformation of cRGD for integrin targeting

The biological activity of cRGD-containing nanodevices is intimately related to the spatial disposition of cRGD ligands and their moieties prior to and after the binding event [61]. Likewise, a successful binding of cRGD to $\alpha_v\beta_3$ integrin depends on the proper conformation acquired by this ligand in solution, and it has been demonstrated that molecular simulations are well-suited to reproduce the pre-organized structure of cyclic RGD peptides [87]. For instance, recent findings have shown that slight changes in the molecular structure of cRGD-based peptides considerably impact their conformational states leading to substantial implications on the binding affinity to $\alpha_v\beta_3$ integrin [88].

To investigate how the ensemble of conformations acquired by the cRGD ligands conjugated to the PEGylated $TiO_2$-based nanodevice is affected by different ligand densities, we analyze the $\Phi$ and $\Psi$ dihedral angles involved in the torsional transitions of the cRGD cycle, as defined in Fig. 2a. Fig. 10 displays the Ramachandran plot for the $\Phi$ and $\Psi$ dihedral angles of cRGD during the MD production phase.

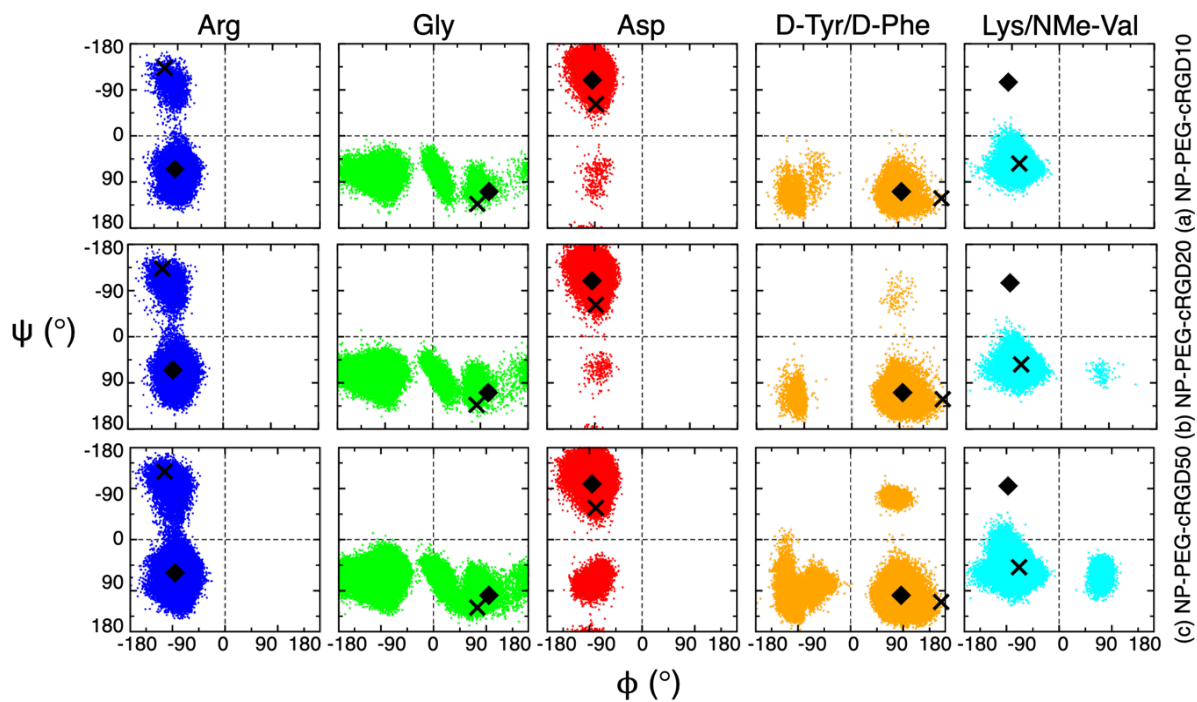

**Fig. 10.** Ramachandran plots of Φ and Ψ dihedral angles of cRGD side chains at (a) 10%, (b) 20% cRGD and (c) 50% cRGD density during the MD production phase. Black crosses identify the dihedral angles of CGT found in the crystal structure of the complex $α_vβ_3$ integrin-CGT (PDB code: 1L5G [60]), while black diamonds are used for dihedral values found by Marelli et al. [63] from MD simulations of a free CGT in water. D-Phe and NMe-Val are CGT amino acids occupying cRGD D-Tyr and Lys place, respectively.

To the best of our knowledge, no crystallographic data is available for the cRGD-$α_vβ_3$ integrin complex to date. Thereby, we use as reference the X-ray structure values of Φ and Ψ dihedral angles of CGT bounded to $α_vβ_3$ [60], being CGT (Arg–Gly–Asp–D-Phe–NMe-Val) a cyclic peptide based on the same RGD sequence and that possesses a similar structure to the one of the cyclic peptide used in this work. As seen in Fig. 10a, b and c, we observe a fair agreement between the Φ and Ψ dihedral angles sampled for Arg, Gly, Asp, and Lys side chains and their equivalent in the X-ray $α_vβ_3$ integrin-CGT complex. Besides, for the D-tyrosine side chain in cRGD, we notice that the Φ and Ψ values found in the crystal structure for D-Phe are only marginally sampled during the MD simulation.

Next, we analyze the main cRGD conformations sampled during the production phase by means of *cluster analysis* (details in Section 2.2.1). We perform the analysis on the most exposed cRGD of each system, that is the one with the longest average distance from the center of the NP. Interestingly, the clusters of cRGD conformation in all the three nanosystems resemble the ones found for the free cRGD in water (Section 3.1.1). The molecular structures for the most relevant conformations are reported in Fig. S5 in Supporting Material.

As previously stated in Section 3.1.1, several experimental works suggest the evaluation of a particular set of geometrical criteria (see Fig. 3c for further details) to assess the potential biological activity of RGD-containing peptides towards integrins [64,65]. Here, we retake the same analysis done for the single cRGD molecule in water (Table 1), further evaluating this set of geometrical parameters for the cRGD-conjugated ligands in the NP-PEG-cRGD10, NP-PEG-cRGD20 and NP-PEG-cRGD50 nanodevices. Table 7 compiles the structural parameters evaluated for CGT X-ray structure, the single cRGD molecule solvated in water and the cRGD ligands.

**Table 7.** Structural parameters of CGT in complex with $α_vβ_3$ integrin [60] and of cRGD averaged over the production phase of MD simulations with their standard deviation. MD predictions for a single cRGD in water and for cRGDs molecules bonded to PEGylated $TiO_2$ nanodevices at 10%, 20% or 50% of cRGD conjugation are reported.

| Structural parameter | CGT (Crystal) [60] | Single cRGD | NP-PEG-cRGD10 | NP-PEG-cRGD20 | NP-PEG-cRGD50 |
|---|---|---|---|---|---|
| $d_1$ (Å) | 13.7 | 11.5 (±1.5) | 11.9 (±1.4) | 11.9 (±1.4) | 11.7 (±0.2) |
| $d_2$ (Å) | 8.9 | 7.6 (±0.6) | 7.7 (±0.6) | 7.6 (±0.6) | 7.7 (±0.1) |
| $θ$ (°) | 136 | 111 (±12) | 108 (±19) | 108 (±20) | 113 (±2) |

Analysis of Table 7 reveals that the $C^ζ_{Arg}$-$C^γ_{Asp}$ $d_1$ and $C^β_{Arg}$-$C^β_{Asp}$ $d_2$ distance values are in fair agreement between the X-ray crystal structure and the cRGD ligands conjugated on the PEGylated $TiO_2$ NP surface. Thereafter, we see no significant changes on these two geometrical parameters for cRGD ligands either as a single molecule free to diffuse in water or conjugated to $PEG_{500}$ polymeric linkers. However, we observe a slight deviation of the $C^β_{Arg}$-$C^α_{Gly}$-$C^β_{Asp}$ $θ$ angle of cRGD compared to the CGT crystal structure in complex with $α_vβ_3$. This deviation is likely linked to the dynamic nature of cRGD in explicit water solvation in contrast to the molecular confinement of CGT into the binding pocket of $α_v β_3$ integrin.

Although the biological activity requires further investigation, we can infer from the analysis above that the cRGD conjugation to PEG linkers has little effect on the conformational states acquired by this ligand. Hence, this result suggests that cRGD is expected to retain its biological activity upon conjugation to PEGylated $TiO_2$ NPs.

## *4. Conclusions*

The MD study on the cRGD-functionalization of PEGylated $TiO_2$ NP models presented in this work provides an atomistic understanding of the underlying reasons why cRGD density is a critical parameter, which profoundly impacts the presentation and effective availability of the targeting ligands on the nanodevice surface, as also suggested by previous experimental investigations on the enhancing strategies of active targeting through the conjugation of PEGylated nanosystems with cRGD ligands at varying densities [7,10,11,81,82]. In particular, our simulations consider three different extents of cRGD-conjugation to PEG chains grafted on a spherical $TiO_2$ NPs of 2.2 nm size. The simulated cRGD densities ranges from 0.2 to 1.0 ligand $nm^{-2}$, broadly comparable to those previously reported in several experimental studies [7,10,11,81,82].

The surface characterization of the cRGD-conjugated PEGylated NP systems has shown that cRGD ligands remain located at the PEG/water interface independently of the ligand density adopted. This observation remarkably agrees with previous experiments by Valencia et al. [11], who found that nearly all cRGD molecules are present on the surface of cRGD-conjugated polymeric NPs, little affected by ligand density changes. Our simulation findings also confirm that cRGD conjugation to polymeric chains slows down the ligand diffusion and has a negligible effect on the cRGD ability to sample optimal conformations, meeting two important requirements for an effective binding activity towards $α_v β_3$ integrin [65,85].

In addition, our simulations predicted a considerable increase in the cRGD-cRGD interactions impairing the ligand's proper solvation and presentation at high ligand density (~1.0 ligand $nm^{-2}$). These simulation results confirm the experimental hypothesis of cRGD overlap and clarify the reduced detection of cRGD on PEGylated NP surfaces at high ligand density regimes [11]. Rather, we found an optimal spatial presentation and negligible over-clustering effects of cRGD ligands at moderate ligand densities (0.2-0.5 ligand nm$^{-2}$). These findings help explain previous experimental data [11,81,82], in which the authors found that NPs conjugated with moderate targeting cRGD densities (0.1-0.7 ligand nm$^{-2}$) provide better targeting efficiency and cellular uptake than higher ligand density regimes (> 1.0 ligand nm$^{-2}$).

From a computational point of view, our study represents a key improvement with respect to the state-of-the-art, since only one computational investigation exists [13], where similar dynamical aspects of PEG-conjugated cRGD on an inorganic surface have been investigated. The advancement with respect to the previous work [13] is related to the fact that we have considered (i) a realistic spherical NP, instead of a flat periodic model, (ii) a semiconducting oxide NP, instead of a simple metallic (Au) surface, and (iii) we have comparatively investigated increasing cRGD densities, instead of full coverage, since the extent of cRGD conjugation is recognized as a critical parameter.

To conclude, in the light of the available experimental data, our theoretical findings provide valuable insights, at an atomistic level, to be exploited in the design of more effective cRGD-based targeting nanosystems. Further development of this work goes in two main directions. On one side, we should learn more about the ligand-receptor binding when the ligand is connected to the NP via PEG chains. On the other, we could enlarge the model, to get closer to real experiments, by increasing the size of the NP and the length of the PEG chains, by playing with the grafting densities and different conformational regimes (e.g., mushroom and brushes regimes). However, such large systems could be investigated only by means of less expensive computational approaches, like coarse-graining methods, paying the price of a reduced chemical accuracy together with the loss of an atomistic description.

## *Conflicts of interest*

The authors declare no conflict of interest.

*Acknowledgments*

The authors are grateful to Lorenzo Ferraro for his technical support and to Daniele Selli and Stefano Motta for useful discussions. The project has received funding from the European Research Council (ERC) under the European Union's HORIZON2020 research and innovation programme (ERC Grant Agreement No [647020]) and from the University of Milano Bicocca FAQC 2020 for the project "Photodynamic therapy for brain tumors by multifunctional particles using in situ Cerenkov and radioluminescence light".

# Supporting Material

# Molecular dynamics simulations of cRGD-conjugated PEGylated TiO$_2$ nanoparticles for targeted photodynamic therapy

Paulo Siani[a], Giulia Frigerio[a], Edoardo Donadoni[a], Cristiana Di Valentin[a,b]

[a]Dipartimento di Scienza dei Materiali, Università di Milano Bicocca,
via R. Cozzi 55, 20125 Milano Italy

[b]BioNanoMedicine Center NANOMIB, University of Milano-Bicocca, Italy

# cRGD and PEG₅₀₀ FF parameters: partial atomic charges

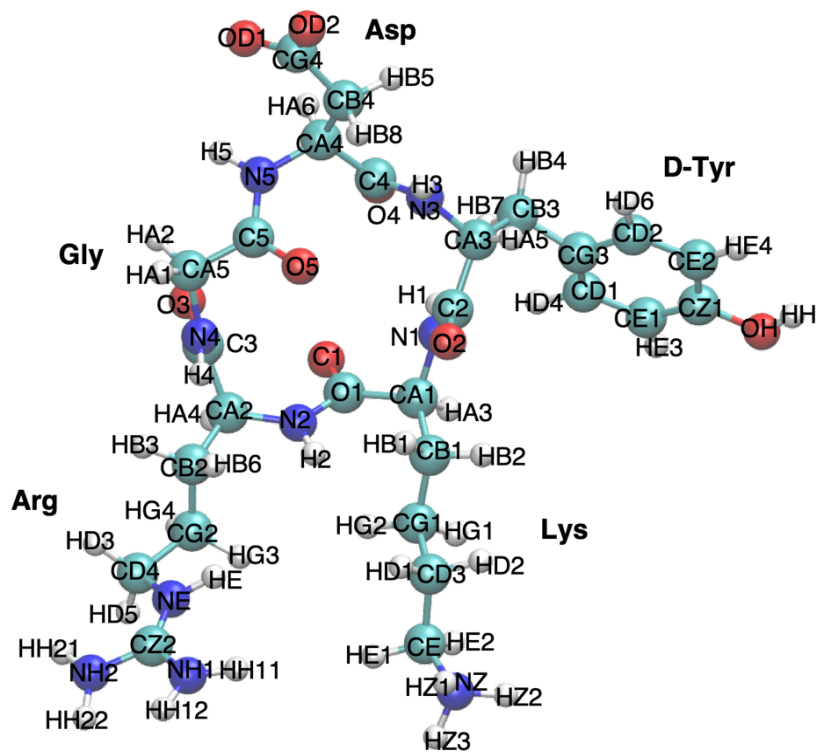

**Fig. S1.** The c(RGDyK) molecule with atom names.

**Table S1.** Partial atomic charges for c(RGDyK) atoms: atom names, CHARMM atom types, partial atomic charges.

| Atom name | CHARMM Atom type | q (e) | Atom name | CHARMM Atom type | q (e) |
|---|---|---|---|---|---|
| **Arg** | | | **D-Tyr** | | |
| C3 | CG2O1 | 0.509 | C2 | CG2O1 | 0.510 |
| O3 | OG2D1 | -0.510 | O2 | OG2D1 | -0.509 |
| CA2 | CG311 | 0.042 | CA3 | CG311 | 0.037 |
| HA4 | HGA1 | 0.090 | HA5 | HGA1 | 0.090 |
| N2 | NG2S1 | -0.436 | N3 | NG2S1 | -0.441 |
| H2 | HGP1 | 0.312 | H3 | HGP1 | 0.312 |
| CB2 | CG321 | -0.183 | CB3 | CG321 | -0.181 |
| HB3 | HGA2 | 0.090 | HB4 | HGA2 | 0.090 |
| HB6 | HGA2 | 0.090 | HB7 | HGA2 | 0.090 |
| CG2 | CG321 | -0.182 | CG3 | CG2R61 | 0.000 |
| HG3 | HGA2 | 0.090 | CD2 | CG2R61 | -0.117 |
| HG4 | HGA2 | 0.090 | HD6 | HGR61 | 0.115 |
| CD4 | CG324 | 0.202 | CD1 | CG2R61 | -0.117 |
| HD3 | HGA2 | 0.090 | HD4 | HGR61 | 0.115 |
| HD5 | HGA2 | 0.090 | CE2 | CG2R61 | -0.111 |
| NE | NG2P1 | -0.702 | HE4 | HGR61 | 0.115 |
| HE | HGP2 | 0.442 | CE1 | CG2R61 | -0.111 |
| CZ2 | CG2N1 | 0.637 | HE3 | HGR61 | 0.115 |
| NH1 | NG2P1 | -0.800 | CZ1 | CG2R61 | 0.108 |
| HH11 | HGP2 | 0.460 | OH | OG311 | -0.529 |
| HH12 | HGP2 | 0.460 | HH | HGP1 | 0.419 |
| NH2 | NG2P1 | -0.800 | **Lys** | | |
| HH21 | HGP2 | 0.460 | C1 | CG2O1 | 0.511 |
| HH22 | HGP2 | 0.460 | O1 | OG2D1 | -0.509 |
| **Gly** | | | CA1 | CG311 | 0.037 |
| C5 | CG2O1 | 0.519 | HA3 | HGA1 | 0.090 |
| O5 | OG2D1 | -0.507 | N1 | NG2S1 | -0.436 |
| CA5 | CG321 | -0.050 | H1 | HGP1 | 0.312 |
| HA1 | HGA2 | 0.090 | CB1 | CG321 | -0.193 |
| HA2 | HGA2 | 0.090 | HB1 | HGA2 | 0.090 |
| N4 | NG2S1 | -0.463 | HB2 | HGA2 | 0.090 |
| H4 | HGP1 | 0.268 | CG1 | CG321 | -0.201 |
| **Asp** | | | HG1 | HGA2 | 0.090 |
| C4 | CG2O1 | 0.510 | HG2 | HGA2 | 0.090 |
| O4 | OG2D1 | -0.509 | CD3 | CG321 | -0.196 |
| CA4 | CG311 | 0.040 | HD1 | HGA2 | 0.090 |
| HA6 | HGA1 | 0.090 | HD2 | HGA2 | 0.090 |
| N5 | NG2S1 | -0.431 | CE | CG324 | 0.168 |
| H5 | HGP1 | 0.354 | HE1 | HGA2 | 0.090 |
| CB4 | CG321 | -0.281 | HE2 | HGA2 | 0.090 |
| HB5 | HGA2 | 0.090 | NZ | NG3P3 | -0.287 |
| HB8 | HGA2 | 0.090 | HZ1 | HGP2 | 0.328 |
| CG4 | CG2O3 | 0.619 | HZ2 | HGP2 | 0.328 |
| OD1 | OG2D2 | -0.760 | HZ3 | HGP2 | 0.328 |
| OD2 | OG2D2 | -0.760 | | | |

**Fig. S2.** PEG$_{500}$-COOH structure with atom names.

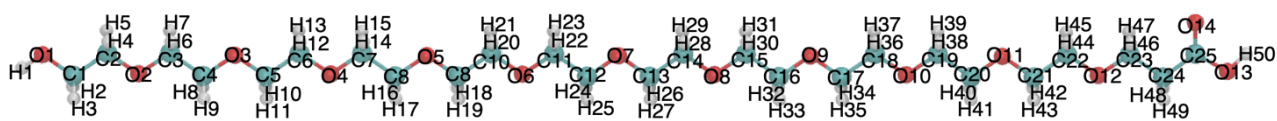

**Table S2.** Partial atomic charges for PEG$_{500}$-COOH atoms: atom names, CHARMM atom types, partial atomic charges.

| Atom name | CHARMM Atom type | q (e) | Atom name | CHARMM Atom type | q (e) |
|---|---|---|---|---|---|
| O1  | OG311 | -0.651 | H7  | HGA2 | 0.090 |
| C1  | CG321 |  0.049 | H8  | HGA2 | 0.090 |
| C2  | CG321 | -0.007 | H9  | HGA2 | 0.090 |
| O2  | OG301 | -0.340 | H10 | HGA2 | 0.090 |
| C3  | CG321 | -0.010 | H11 | HGA2 | 0.090 |
| C4  | CG321 | -0.010 | H12 | HGA2 | 0.090 |
| O3  | OG301 | -0.340 | H13 | HGA2 | 0.090 |
| C5  | CG321 | -0.010 | H14 | HGA2 | 0.090 |
| C6  | CG321 | -0.010 | H15 | HGA2 | 0.090 |
| O4  | OG301 | -0.340 | H16 | HGA2 | 0.090 |
| C7  | CG321 | -0.010 | H17 | HGA2 | 0.090 |
| C8  | CG321 | -0.010 | H18 | HGA2 | 0.090 |
| O5  | OG301 | -0.340 | H19 | HGA2 | 0.090 |
| C9  | CG321 | -0.010 | H20 | HGA2 | 0.090 |
| C10 | CG321 | -0.010 | H21 | HGA2 | 0.090 |
| O6  | OG301 | -0.340 | H22 | HGA2 | 0.090 |
| C11 | CG321 | -0.010 | H23 | HGA2 | 0.090 |
| C12 | CG321 | -0.010 | H24 | HGA2 | 0.090 |
| O7  | OG301 | -0.340 | H25 | HGA2 | 0.090 |
| C13 | CG321 | -0.010 | H26 | HGA2 | 0.090 |
| C14 | CG321 | -0.010 | H27 | HGA2 | 0.090 |
| O8  | OG301 | -0.340 | H28 | HGA2 | 0.090 |
| C15 | CG321 | -0.010 | H29 | HGA2 | 0.090 |
| C16 | CG321 | -0.010 | H30 | HGA2 | 0.090 |
| O9  | OG301 | -0.340 | H31 | HGA2 | 0.090 |
| C17 | CG321 | -0.010 | H32 | HGA2 | 0.090 |
| C18 | CG321 | -0.010 | H33 | HGA2 | 0.090 |
| O10 | OG301 | -0.340 | H34 | HGA2 | 0.090 |
| C19 | CG321 | -0.010 | H35 | HGA2 | 0.090 |
| C20 | CG321 | -0.010 | H36 | HGA2 | 0.090 |
| O11 | OG301 | -0.340 | H37 | HGA2 | 0.090 |
| C21 | CG321 | -0.010 | H38 | HGA2 | 0.090 |
| C22 | CG321 | -0.010 | H39 | HGA2 | 0.090 |
| O12 | OG301 | -0.340 | H40 | HGA2 | 0.090 |
| C23 | CG321 | -0.005 | H41 | HGA2 | 0.090 |
| C24 | CG321 | -0.219 | H42 | HGA2 | 0.090 |
| C25 | CG2O2 |  0.769 | H43 | HGA2 | 0.090 |
| O13 | OG311 | -0.603 | H44 | HGA2 | 0.090 |
| O14 | OG2D1 | -0.562 | H45 | HGA2 | 0.090 |
| H1  | HGP1  |  0.419 | H46 | HGA2 | 0.090 |
| H2  | HGA2  |  0.090 | H47 | HGA2 | 0.090 |
| H3  | HGA2  |  0.090 | H48 | HGA2 | 0.090 |
| H4  | HGA2  |  0.090 | H49 | HGA2 | 0.090 |
| H5  | HGA2  |  0.090 | H50 | HGP1 | 0.430 |
| H6  | HGA2  |  0.090 |     |      |       |

**Simulated annealing of cRGD**

The same simulation protocol described in the main text was used both for the all-*trans* conformation Fig. S3 (a) and for the *cis-trans* one in Fig. S3 (b), where G-D and K-R bonds are *cis* and the other ones *trans*. After a production phase of 100 ns was run, the two c(RGDyK) conformations were subjected to the following simulated annealing procedure, which allows for a minimum global search: first, the temperature is heated up to 700 K at a rate of 2 K/step; second, the system is let evolve in time for 50 ns; then it is cooled down to the original temperature at a rate of 2 K/step and finally allowed to evolve for another 50 ns. For both the all-*trans* conformation and the *cis-trans* one the temperature of 700 K allowed us to observe amide bond *cis-trans* isomerization. In particular, Gly-Asp amide bond isomerized more often than other ones, but eventually the structures ended up in an all-*trans* conformation during the cooling down phase. Thus, the *all*-trans conformation was adopted for this work.

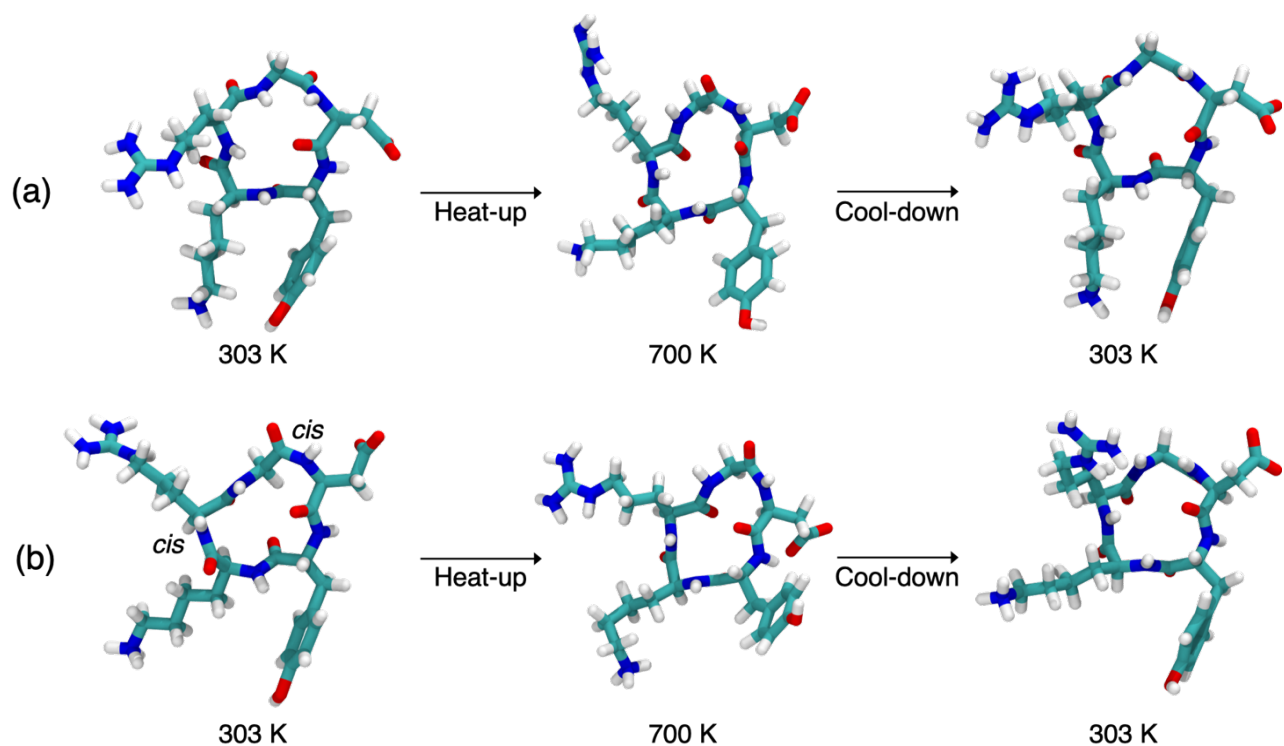

**Fig. S3.** Starting-point geometry, an intermediate structure and the last snapshot of the simulated annealing procedure for the all-*trans* structure (a) and the *cis-trans* structure (b) of cRGD, respectively.

**PEG$_{500}$ dihedral angle distribution**

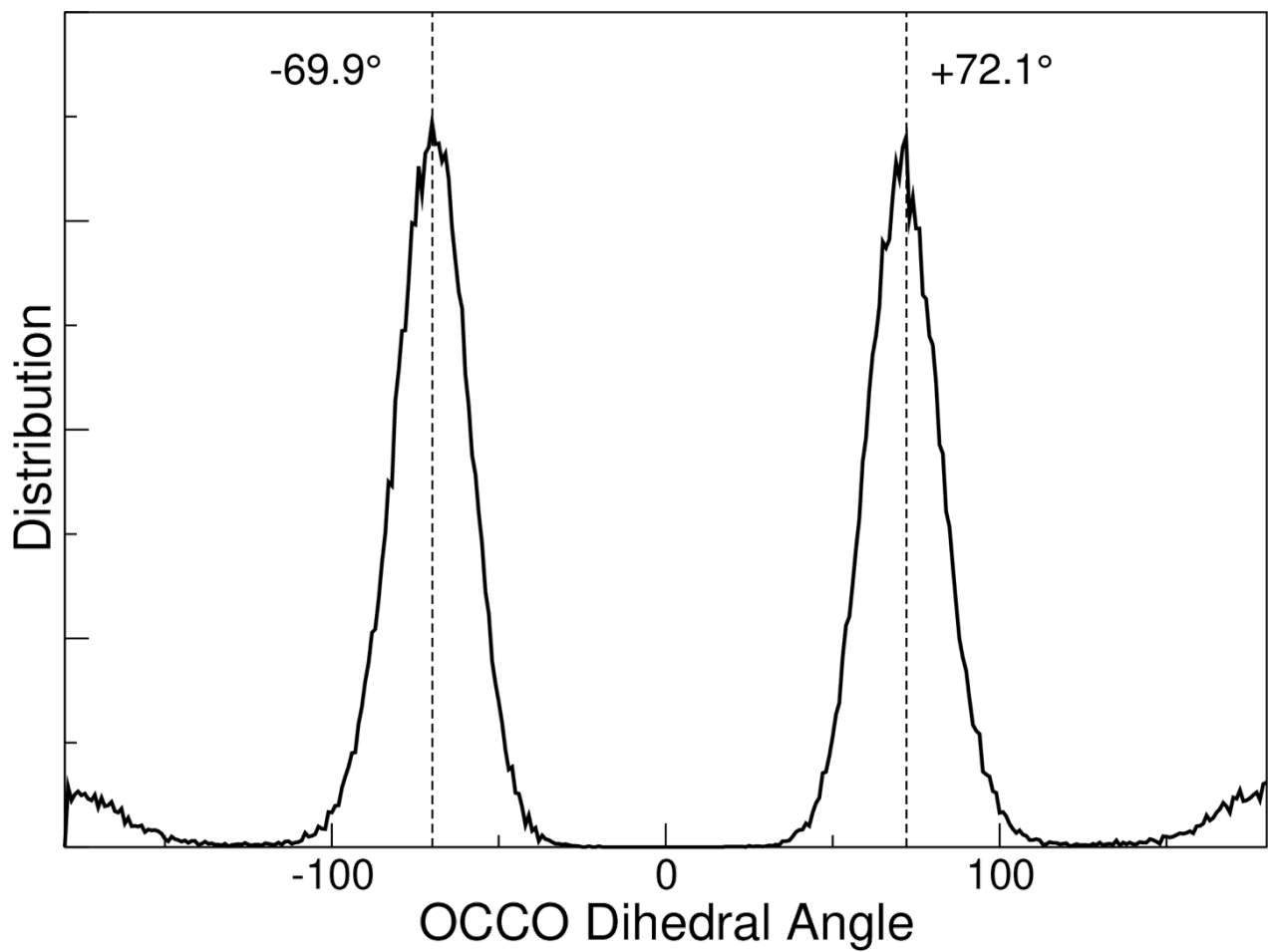

**Fig. S4.** Polymer dihedral distribution of a single PEG$_{500}$ chain in water.

## Self-diffusion coefficients

**Table S3.** Self-diffusion coefficients (*D*) of single cRGD in water and bonded to PEGylated TiO$_2$ nanodevices at 10%, 20% and 50% of cRGD conjugation with their standard deviation.

| Self-diffusion coefficients *D* (m$^2$ s$^{-1}$) | |
|---|---|
| **Single c(RGDyK)** | 1.24·10$^{-8}$ (±2·10$^{-10}$) |
| **NP-PEG-cRGD10** | 1.19·10$^{-9}$ (±2·10$^{-11}$) |
| **NP-PEG-cRGD20** | 7.42·10$^{-10}$ (±7·10$^{-12}$) |
| **NP-PEG-cRGD50** | 6.82·10$^{-10}$ (±4·10$^{-12}$) |

**Cluster analysis for NP-PEG-cRGD10, NP-PEG-cRGD20 and NP-PEG-cRGD50**

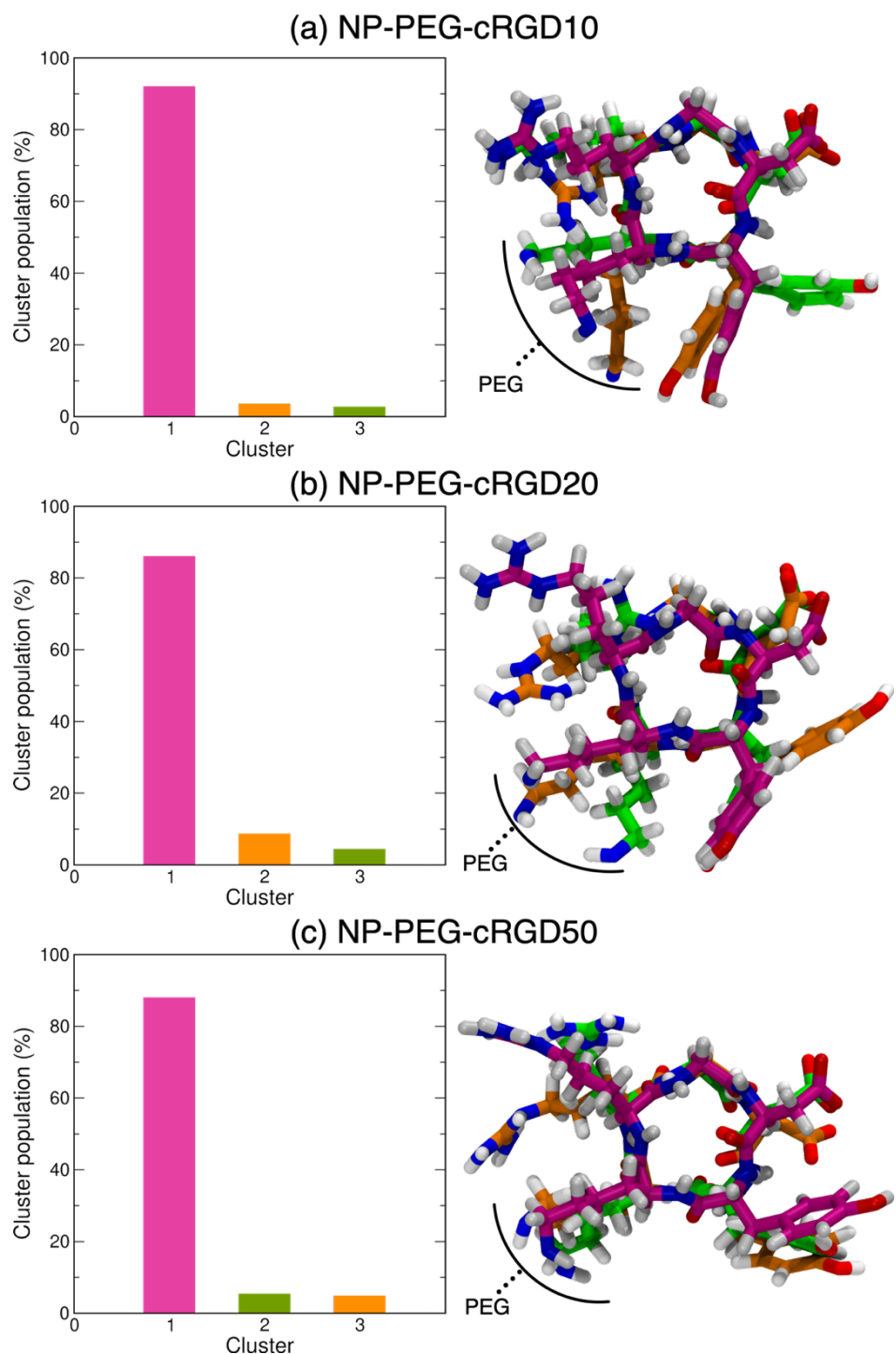

**Fig. S5.** Cluster occupancy and average structures of the first three clusters in order of population for (a) NP-PEG-cRGD10, (b) NP-PEG-cRGD20 and (c) NP-PEG-cRGD50 nanosystems. The analysis is performed on the most exposed cRGD. The other 5 clusters represent less than 1% of the conformations adopted during the production phase. Carbon atoms are of the same color of the cluster, while oxygen, nitrogen, and hydrogen atoms are shown in red, blue, and white, respectively.